\pdfoutput=1
\documentclass[12pt]{article}

\usepackage[utf8]{inputenc}
\usepackage[margin=1in]{geometry}
\usepackage{authblk}
\usepackage{abstract}
\usepackage{setspace}
\usepackage{parskip} 
\usepackage{array}
\usepackage{ragged2e}
\usepackage{booktabs}
\usepackage{rotating}
\usepackage{enumitem}
\usepackage{graphicx}
\usepackage[tagged]{accessibility}
\usepackage{needspace}
\usepackage{etoolbox}

\preto\section{\needspace{4\baselineskip}}
\preto\subsection{\needspace{4\baselineskip}}
\preto\subsubsection{\needspace{3\baselineskip}}

\makeatletter
\renewcommand\@biblabel[1]{}
\makeatother

\title{\textbf{The Digital Gorilla: Rebalancing Power in the Age of AI}}

\author{
M. Alejandra Parra-Orlandoni\thanks{Senior Fellow, Mossavar-Rahmani Center for Business and Government, Harvard Kennedy School.}, 
Roxanne A. Schnyder\thanks{L.L.M. candidate, Harvard Law School.}, 
Christopher J. Mallet\thanks{M.P.A. candidate, Harvard Kennedy School.}
}

\date{4 February 2026}

\begin{document}

\maketitle

\begin{abstract}
Contemporary artificial intelligence (AI) policy suffers from a basic categorical error. Existing frameworks rely on analogizing AI to inherited technology types -- such as products, platforms, or infrastructure -- and in doing so generate overlapping, often contradictory governance regimes. This ``analogy trap'' obscures a fundamental transformation: certain advanced AI systems no longer function solely as instruments through which existing institutions exercise power, but as de facto centers of power that shape information, coordinate behavior, and structure social and economic realities at scale. This article offers a new conceptual foundation for AI governance by treating such systems as a fourth societal actor -- what we term the ``Digital Gorilla'' -- alongside People, the State, and Enterprises. It develops a Four Societal Actors framework that maps how power flows among these actors across five power modalities (economic, epistemic, narrative, authoritative, physical) and uses this map to diagnose where AI capabilities disturb established allocations of authority, concentrate power, or erode accountability. Drawing on constitutional principles of separated powers and federalism, the article advances a federalized, polycentric governance architecture and institutionalizes dynamic checks and balances among the four actors, rather than today's more reactive and compliance-driven approaches. Reframing AI governance in this way shifts the inquiry from how to control a risky technology to how to design institutions capable of accommodating these increasingly powerful and autonomous digital systems without sacrificing democratic legitimacy, the rule of law, or the production of public goods, and it recasts familiar debates in administrative, constitutional, and corporate law as questions of power allocation in a four-actor system.
\end{abstract}

\vspace{1em}

\noindent\textbf{Keywords:} AI governance; AI policy; AI regulation; algorithmic governance; constitutional design; institutional design; polycentric governance; democratic legitimacy; algorithmic accountability; AI as a societal actor

\clearpage

\section{Introduction: The Arrival of the Digital Gorilla}

Public discourse about AI reflects a deep conceptual instability. As AI systems routinely write, design, analyze, predict, and coordinate at levels that rival human capability, the familiar lenses through which we view technology have become inadequate. Popular anxiety captures this dislocation -- if AI paints better, writes better, governs better, why do humans still matter? -- but the underlying issue is more classificatory than existential. Many emerging AI systems are no longer experienced purely as sociotechnical instruments but as actors in society that actively shape social, economic, and political life.

Now imagine a room where policymakers, technologists, and citizens debate the future of artificial intelligence (AI). All eyes focus on the rules, regulations, and governance frameworks, but no one acknowledges what we have termed the ``Digital Gorilla'' sitting in the corner, reshaping the conversation by its mere presence. This, we argue, is the state of AI policy\footnote{For convenience, we use the shorthand ``AI policy'' and ``AI governance'' throughout this document to refer to the broad collection of laws, regulations, guidance, norms, ethics and other formal and informal mechanisms meant to manage how AI systems are developed, deployed, and used.} today: we legislate and regulate as if AI were a tool, when it increasingly functions as an actor in its own right.

This shift has historical antecedents. When the 1968 film \textit{2001: A Space Odyssey} introduced HAL 9000, a machine capable of autonomous decision-making, the film's technical advisor, mathematician I. J. Good, had already theorized an ``intelligence explosion'' of recursive self-improvement beyond human comprehension (Good 1966). While Good's formulation was speculative, contemporary AI systems already display properties he identified, exercising forms of de facto agency. These systems screen job applicants, determine loan approvals, curate global information flows, coordinate logistics networks, author original content, and execute financial trades. Unlike traditional software, they do so with minimal human intermediation and generate effects traditionally associated with organized human institutions.

We therefore argue that AI should be conceptualized as a fourth societal actor, alongside People, the State, and Enterprises, as these terms are defined below in Section 2.2. This parallels the treatment of states and enterprises as institutional actors, not because they possess human qualities, but because their structural position and societal impacts warrant distinct policy approaches.

Characterizing AI as a societal actor is a functional, not ontological, claim. It does not attribute personhood to AI systems, nor does it require resolving questions about AI's consciousness or moral status (Butlin et al. 2023; Schwitzgebel \& Garza 2015) to recognize that AI systems perform in ways that exceed traditional tool- or infrastructure-like behaviors. Rather, the claim reflects a socio-institutional reality that contemporary AI systems increasingly possess decision-making power over consequential outcomes, the capacity for large-scale coordination and implementation, and the ability to structure preferences, constrain choices, and distribute resources.

The contribution of this article is threefold. First, it dissects the shortcomings of prevailing AI policy paradigms and proposes, as a replacement, a conceptual model that treats a subset of AI systems as societal actors (Section 2). Second, it maps the resulting four-actor power dynamics, showing how AI alters checks and balances among People, the State, and Enterprises and how policy can recalibrate power dynamics at a system level (Section 3). Third, it derives implications for institutional design, regulatory strategy, and governance, providing a coherent framework for governing AI systems whose capabilities no longer merely augment, but intersect with, core functions of societal organization (Section 4).

Although the arguments presented are largely conceptual, they carry concrete doctrinal implications. Treating a subset of AI systems as societal actors clarifies problems that existing doctrines only partially address. In public law, AI-driven systems increasingly mediate policing, access to benefits, and core civil liberties, straining due process, equal protection, and nondelegation doctrines that presuppose human decision-makers. In administrative law, agencies increasingly rely on AI to generate, interpret, and execute policy, yet standards of delegation, reason-giving, and judicial review remain tied to human officials. In private law, corporate and fiduciary structures authorize firms to deploy AI at scale without a clear account of where responsibility lies when AI systems function as de facto decision-makers. By reconceptualizing AI as a societal actor and mapping power dynamics across actors, this article offers a jurisprudential framework for understanding doctrinal debates as questions of power allocation and institutional design rather than as narrower questions of risk.

Finally, the analysis and proposals in this article rely on the institutional substrates of liberal-democratic states with market-oriented economies. While the revelation of the Digital Gorilla is a global phenomenon, the mechanisms of federalism, checks and balances, and civil liberties presuppose a governance structure capable of distributing power. How these dynamics play out in authoritarian, state-capitalist, or other political-economic systems is a critical but separate inquiry.

\section{Reframing AI as a Societal Actor: The Four Societal Actors Model}

\subsection{The Analogy Trap: Why Existing AI Policy Models Fail}

Current AI policy debates oscillate between two ideologically entrenched fears: that regulation will stifle innovation and that insufficient oversight will expose the public to catastrophic harms. This mirrors past debates about technologies such as biotechnology, nuclear energy, and the internet, in which policymakers sought to balance economic growth against public risk under conditions of technological uncertainty.

For example, debates over pharmaceutical regulation since the 1962 Kefauver-Harris Amendments produced recurring claims that stricter Food and Drug Administration safety and efficacy review created a ``drug lag,'' depriving patients of beneficial new therapies in the name of safety (Drug Amendments of 1962; Schifrin \& Tayan 1977). In the nuclear domain, early civilian nuclear power policy, culminating in the Atomic Energy Act of 1954 and the Atoms for Peace program, likewise attempted to reconcile the promise of cheap, abundant energy with fears of reactor accidents and proliferation (Atomic Energy Act of 1954; Hewlett \& Holl 1989). More recently, the net-neutrality fights over the Federal Communication Commission's Open Internet Order pitted advocates of nondiscrimination rules to protect users and free expression against critics warning that such regulation would hobble broadband innovation (Federal Communications Commission 2010; Parker 2015).

Likewise, governments today are pursuing deregulatory strategies for AI in the name of innovation and geopolitical competitiveness (European Commission 2025; Bordelon \& Mui 2025) and industry actors are warning that compliance costs will drive development to less restrictive jurisdictions (Roberts 2025; Johnson 2025; Draghi 2024). At the same time, civil society organizations are documenting accumulating social harms of AI, including bias, opacity, and exclusion (OECD AI Incidents Monitor; MIT AI Risk Repository), while prominent AI researchers, such as Geoffrey Hinton and Yoshua Bengio, publicly issue warnings about high-impact risks associated with advanced AI systems (Hinton 2024; Bengio 2024).

Despite the evident complexity and multidimensionality of these issues, mainstream narratives about AI largely frame policy intervention as a zero-sum game in which safety necessarily sacrifices innovation or vice versa. Consider, for example, the Trump administration's positioning on AI regulation, summarily captured by U.S. Vice President Vance's declaration at the 2025 Paris AI Action Summit that ``The A.I. future is not going to be won by hand-wringing about safety'' (Sanger 2025). In contrast, Center for AI and Digital Policy President Merve Hickok stated that the European Commission's proposed Digital Omnibus ``will let loose unsafe AI systems in the EU that will threaten public safety and fundamental rights'' (Jahangir 2025).

This impasse is not simply a matter of competing political or economic interests. It reflects a deeper conceptual error, and we use the well-known duck-rabbit illusion to illustrate it (Wittgenstein 1953/1967). Viewed as a ``duck,'' AI appears as a conventional, human-deployed technological instrument, whose risks can be managed through incremental safeguards, ex post liability, and sector-specific regulation. Reoriented as the ``rabbit,'' the same AI systems come into focus as actors that structure societal outcomes in ways that are neither fully predictable nor fully attributable to any single human actor, exposing the shortcomings of policies grounded in instrumental analogies. The failure of existing AI policy models, this article argues, lies in their inability to see beyond the duck.

\subsubsection{Here Is the Duck: The Misclassification Problem}

Policymakers remain ensnared in what may be termed an analogy trap: the instinct to map AI onto inherited technology categories like ``software,'' ``infrastructure,'' ``platform,'' or ``general-purpose technology'' to make AI legible within existing governance frameworks. These classifications provide juridical shortcuts, allowing regulators to extend familiar doctrines -- product safety (for example, through the EU AI Act and proposed AI Liability Directive), data protection (such as the General Data Protection Regulation and related case law such as Schufa Holding AG 2023), and intermediary liability (for example, the Digital Services Act and Communications Decency Act \S~230) -- to AI, even as the distinctive scale, opacity, and adaptivity of contemporary AI systems stretch those frameworks beyond their limits, precipitating doctrinal cracks.

A parallel pattern has emerged in legal debates over digital currencies, where courts and regulators have struggled to classify cryptocurrencies as money, property, securities, or commodities. Each analogy captures certain functional aspects while obscuring others. In the United States, for example, Bitcoin and other digital assets are treated simultaneously as property for federal tax purposes, commodities subject to Commodity Futures Trading Commission jurisdiction, and securities when they satisfy the investment-contract test (Internal Revenue Service 2014; CFTC v. McDonnell 2018). In the United Kingdom, courts and regulators likewise oscillated among property, financial-instrument, and unregulated-token characterizations before recognizing cryptoassets as a novel form of property and extending regulatory oversight (AA v. Persons Unknown 2019; Law Commission 2023).

These borrowed analogies have played a significant role in structuring contemporary AI policy across jurisdictions. As elaborated below, they have architected the European Union's AI Act, which alternately frames AI systems as products subject to conformity assessment and ex ante risk-management obligations, as infrastructure embedded within regulated sectors through application-based risk tiers, and as intermediary systems subject to transparency and information-disclosure duties (Regulation (EU) 2024/1689). They have likewise shaped U.S. debates over foundation-model governance, where such models are conceptualized as general-purpose technologies warranting downstream, sector-specific oversight, as critical infrastructure requiring pre-deployment evaluation, and as platform-like intermediaries implicating competition and consumer-protection law (National Institute of Standards and Technology 2023; Exec. Order 14110; Khan 2023). In China, national AI strategy similarly reflects shifting analogical frames, treating AI at once as a strategic industrial technology central to economic modernization, as information infrastructure subject to security and data-localization obligations, and as a content intermediary governed through platform-style regulation (State Council of the People's Republic of China 2017; Cybersecurity Law (China); Data Security Law (China); Personal Information Protection Law (China); Algorithmic Recommendation Provisions (China); Interim Measures on Generative AI Services (China)).

Each of these analogies reflects a partial understanding of AI systems and illuminates some juridical perspectives while obscuring others, producing what philosopher Ludwig Wittgenstein described as an aspect shift: AI appears alternately as a tool subject to product liability, as infrastructure subject to safety regulation, or as an intermediary subject to platform governance (Wittgenstein 1953/1967). What is missing is a unifying jurisprudential account of how these systems exercise influence across institutional domains. As a result, AI governance remains trapped in this unstable aspect shift, oscillating between incompatible legal frames rather than resolving the underlying conceptual tensions.

As anomalies in AI governance approaches accumulate, they signal a Kuhnian strain (Kuhn 2012) in the prevailing paradigm, where governing metaphors no longer fit the empirical reality (Birhane 2022). This manifests as a networked-governance failure. As policymakers legislate and regulate through partial analogies, interventions fragment across legal domains (data protection, product safety, consumer protection, competition, and national security) without a shared account of causal dependencies, leverage points, or responsibility allocation. Empirical mapping efforts, such as the MIT AI Governance Mapping project (MIT AI Risk Repository), reinforce this diagnosis, finding that regulatory attention is disproportionately concentrated late in the AI lifecycle (deployment, operation, monitoring) and framed at broad technical levels (``AI systems/models''), such that governance often arrives only after design and architecture choices have already propagated system-wide effects that are difficult to unwind through ex post legal controls (Easley \& Kleinberg 2010).

The consequences of this doctrinal exhaustion are visible across regulatory practice. Platform and social media analogies, for example, have extended content moderation and data protection frameworks to AI systems, applying instruments such as General Data Protection Regulation to algorithmic decision-making and treating AI-generated content as a variant of user-generated content (Regulation (EU) 2016/679; Regulation (EU) 2022/2065). This lens foregrounds individual rights (transparency, contestability, redress) but struggles to address aggregate epistemic effects or structural power asymmetries that emerge from AI systems operating at scale (Ananny \& Crawford 2018, Zuboff 2019). By construing harms as discrete violations of individual rights and channeling remedies through subject-centered claims, it leaves little room to contest how AI systems reorganize public discourse, concentrate ``data power,'' or reshape the conditions for democratic knowledge production (Cohen 2019, Lynskey 2019).

General-purpose technology analogies have driven tiered risk-based regulation, most prominently in the European Union AI Act's classification of AI systems by application context (unacceptable, high-risk, limited-risk, minimal-risk) rather than by capability (Regulation (EU) 2024/1689). This approach frames AI as enabling infrastructure, like electricity or steam power, for which risks depend on deployment context. This leads to category-specific safety requirements and conformity assessments modeled on product safety regimes (Veale \& Borgesius 2021). As a result, regulatory scrutiny tends to focus on specific deployment contexts and product uses, after capability and architecture choices have already been set, leaving questions of upstream power concentration and infrastructural control comparatively underexamined. Some critiques of the AI Act underscore this concern by arguing that it borrows heavily from product safety legislation and focuses obligations at market entry while inadequately addressing lifecycle-wide impacts and the realities of general purpose AI and ``AI as a service'' offerings (Ada Lovelace Institute 2022).

Critical infrastructure comparisons have shaped proposals for AI system testing and certification regimes analogous to those governing nuclear facilities, aviation, and pharmaceuticals, emphasizing pre-deployment safety validation, ongoing monitoring, and incident reporting (Falco et al. 2021; Schuett 2023). Institutions such as the United Kingdom's AI Security Institute and the U.S. National Institute of Standards and Technology exemplify this approach, developing evaluation frameworks and ``safety cases'' borrowed from high-reliability engineering (U.K. AI Security Institute 2024; National Institute of Standards \& Technology 2023). While valuable for managing certain categories of risk, these frameworks again presuppose an object of regulation whose risks can be specified and contained ex ante, an assumption that sits uneasily with adaptive, networked, and generative systems.

Finally, industrial and technological revolution framings have positioned AI as a driver of productivity growth and economic competition, resulting in national strategies prioritizing computer infrastructure investment, talent acquisition, and maintaining technological leadership (Lee 2018). The U.S. CHIPS and Science Act, China's New Generation AI Development Plan, and the European Union's AI competitiveness initiatives all reflect this orientation, treating AI governance primarily as industrial policy but neglect individual and community-level impacts (CHIPS and Science Act 2022; State Council of the People's Republic of China 2017; Draghi 2024). By construing AI chiefly as an engine of growth and geopolitical advantage, these economic framings further entrench an instrumental conception of AI and reinforce regulatory approaches that prioritize expansion and control over structural accountability.

Each analogy generates its own governance logic, and the result is an incoherent patchwork of regulatory approaches. Such fragmentation is inevitable when law operates without a coherent theory of the object it purports to govern. Privacy provides an example in which conceptual under-theorization has resulted in an internally inconsistent body of doctrine and regulation (Solove 2006; Cohen 2013).

Such misclassifications of AI matter not only because they shape risk assessments, jurisdiction, and harm recognition, but because they determine when (and whether) concentrated algorithmic power becomes visible to law at all, often only after design choices have already reshaped social and institutional order. Unlike earlier technologies, which primarily amplified the power of existing actors (the printing press amplified speech; electricity powered industry; the internet globalized connectivity), contemporary AI systems operate as power-wielding entities in their own right (Lukes 2005). The treatment of AI as a technology within inherited policy paradigms obscures this critical shift from influence to direct exertion of power (Lessig 1999) and helps explain the misaligned policy environment now emerging.

\subsubsection{From Misclassification to Policy Incoherence}

To be fair, analogy-based reasoning is a normal mode of institutional learning. Policymakers often, and successfully, draw on established frameworks rather than reinvent governance from scratch (Lakoff \& Johnson 1980; Kahneman 2011). But AI's distinctiveness strains these inherited models (Stone 2012). Risk-based frameworks struggle with systems that evolve unpredictably; rights-based governance focuses on individual harms while missing how AI systems reorganize decision-making at scale; innovation-oriented policies optimized for economic growth sidestep concerns about epistemic authority, information asymmetry, and narrative control (see Section 2.1.1).

These tensions manifest in policy contradictions. For instance, export controls designed to limit adversaries' access to advanced compute (Bureau of Industry and Security 2022; Bureau of Industry and Security 2023) coexist uneasily with domestic safety regimes that fail to mitigate catastrophic risks arising from those same capabilities (Shavit 2023). Moreover, many of these policies operate in isolation, creating irreconcilable incentives and lacking workable framework for adjudicating the tensions (Heng 2025).

As the faulty metaphors lose explanatory power, the AI policy landscape exhibits the features of a paradigm crisis (Kuhn 2012). The categories used to govern AI no longer match how systems are built or deployed, and doctrinal tools pull in conflicting directions. Acknowledging AI's distinctiveness is necessary but insufficient; cataloging differences does not yield a workable model for how AI should fit within existing governance architectures. Instead, AI requires a more robust conceptual framework that renders it legible as an emerging participant in social, economic, and political systems. Only with such a framework can law treat AI as an object of jurisprudential analysis, rather than as a target of primarily technology policy.

Critically, conceiving of AI as a societal actor does not anthropomorphize it; rather, it reflects the reality that AI systems are increasingly embedded in institutional processes, shaping power flows alongside states, firms, and individuals. In a sense, AI has become the proverbial 800-pound gorilla -- impossible to ignore and fundamentally reshaping the space it occupies. The analogy trap persists because we mischaracterize what AI has become. Section 2.2 demonstrates why contemporary AI systems warrant recognition as societal actors rather than mere tools.

\subsection{Here Is the Rabbit: AI as a Societal Actor}

We now return to the duck-rabbit illusion: with a slight shift in perspective, the same lines appear now as a duck, now as a rabbit. This section argues that AI policy and governance frameworks are caught in a similar perceptual trap, seeing only the ``duck'' (AI as technology) when they also need to see the ``rabbit'' (AI as actor). What appears from one angle like sophisticated software begins, from another, to resemble an organized institutional actor -- something whose operations other actors must anticipate, negotiate with, constrain, and so on. Put differently, the same underlying systems can be understood either as tools or as counterparts in institutional life, depending on the lens we choose. This is the ``rabbit'' in the illusion: AI appears to be a tool only as long as we insist on seeing it as one.

In this article, ``AI-as-actor'' serves a diagnostic frame, not a metaphysical claim. It helps surface where AI systems behave like strategic counterparts in institutional settings, reshaping incentives, dependencies, and constraints across existing actors. The point is not to reclassify AI as a moral agent but to make visible the coordination and legitimacy challenges that emerge when decision-making is partially delegated to systems that operate at speed, scale, and opacity. Once we adjust the frame, AI's operational footprint suggests a qualitatively distinct new societal actor emerging alongside People, the State, and Enterprises. See Figs. 1a, 1b.

\begin{figure}[h]
\centering
\begin{minipage}{0.45\textwidth}
\centering
\includegraphics[width=\textwidth, alt={A diagram displaying three round nodes (Enterprises, State, and People) arranged in a triangle, with each node connected to the other nodes via two-way arrows, and where the Enterprises and State nodes are lighter in color with dashed outlines and the People node is darker in color with a solid outline. The three nodes are set on top of a background image of Wittgenstein’s “duck-rabbit” illusion, where the same line drawing seen from one aspect looks like a duck and from another aspect looks like a rabbit.}]{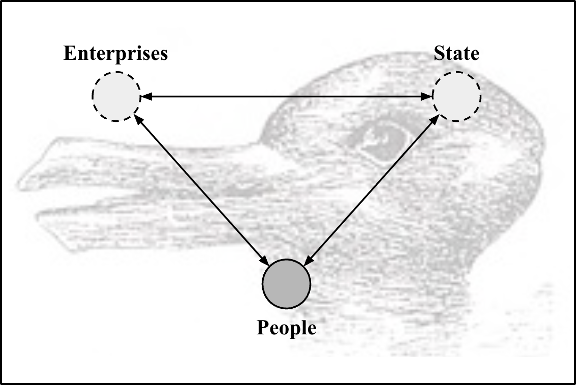}
\caption{Three-actor model (``the duck''): People -- State -- Enterprises}
\label{fig:1}
\end{minipage}
\hfill
\begin{minipage}{0.45\textwidth}
\centering
\includegraphics[width=\textwidth, alt={A diagram displaying four round nodes (Enterprises, State, People, and AI) where the Enterprises, State, and AI notes are arranged in a triangle and the People node is in the center, with each node connected to the other nodes via two-way arrows, and where the Enterprises, State, and AI nodes are lighter in color with dashed outlines and the People node is darker in color with a solid outline. The four nodes are set on top of a background image of Wittgenstein’s “duck-rabbit” illusion, where the same line drawing seen from one aspect looks like a duck and from another aspect looks like a rabbit.}]{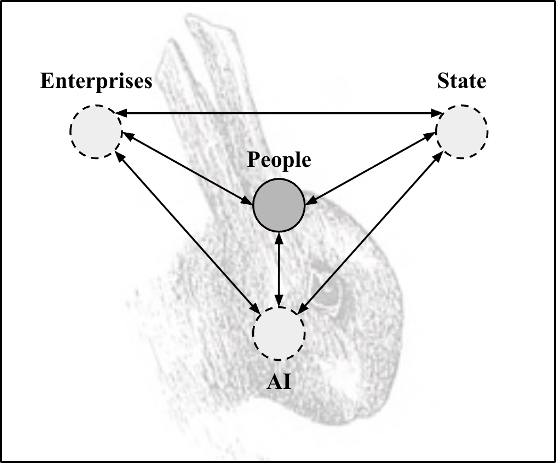}
\caption{Four-actor model (``the rabbit''): People -- State -- Enterprises -- AI}
\label{fig:2}
\end{minipage}
\end{figure}

In these actor models, People represent individuals and communities, the primordial source of human agency from which other institutional actors arise. The State encompasses governmental institutions. Enterprises include corporations, financial institutions, and other centers of private economic power. Institutions such as universities, non-governmental organizations, and international bodies align with one of the primary actors based on critical factors like funding and accountability structures. For example, a state-funded research lab may act as an extension of State power, while a private university might function within the Enterprise logic. Finally, technologies -- represented by the arrows in the figures -- function as tools, media, or environments that facilitate exchange between actors or provide infrastructure through which they can act. In the Four Societal Actors model, a subset of AI systems ceases to merely connect the existing actors but warrants representation as actors in their own right.

\subsubsection{Defining `Societal Actor'}

To govern AI effectively, policymakers must be able to distinguish between AI systems as technologies and AI systems as actors. Although defining ``actorhood'' philosophically and then checking whether an AI systems qualifies seems a promising direction, it is not the approach we take here. For those interested in deeper engagement regarding fundamental questions about the evolving definition of actorhood itself, consider actor-network theory, which treats agency as distributed across human and non-human elements (Latour 2005); algorithmic governance scholarship, which examines how rule systems operate through computational processes (Katzenbach \& Ulbricht 2019); responsibility-gap analysis, which explores accountability when autonomous systems make consequential decisions (Matthias 2004; Santoni de Sio \& Mecacci 2021); and platform sovereignty arguments, which analyze how digital infrastructures exercise quasi-governmental functions (Pohle \& Thiel, 2020).

For the purposes of this article, the functional observation -- that AI systems are already treated as actors by other institutional actors -- provides sufficient foundation for analysis. Thus, we approach the claim that AI functions as a societal actor empirically: when do we, in practice, treat an entity as an actor rather than a tool (Bijker, Hughes \& Pinch 1987)?

Historically, we came to treat corporations and the state as actors not because we believed they possessed human agency, a soul, or some other intrinsic personhood, but because institutional life made it practical to do so. Other actors, like individuals and public authorities, were already responding to corporate entities as if they were actors, negotiating with them, regulating them, holding them accountable, and being constrained by them (Dewey 1926). Legal doctrines such as corporate personhood therefore emerged, recognizing entities can hold assets, enter contracts, sue and be sued, and persist beyond any individual member (Trustees of Dartmouth College v. Woodward 1819; Bank of the United States v. Deveaux 1809; Santa Clara County. v. South Pacific Railroad Co. 1886). In parallel, modern statecraft treats governments as unitary actors for purposes of diplomacy, bargaining and coercion, even though their ``decisions'' emerge from complex internal coalitions and bureaucratic processes (United States v. Curtiss-Wright Exp. Corp. 1936; Zivotofsky ex rel. Zivotofsky v. Kerry 2015; Banco Nacional de Cuba v. Sabbatino 1964). In both cases, the legal and administrative apparatus stabilizes responsibility, creates identifiable decision points, and enables other actors to negotiate with, constrain, and hold these entities accountable, even absent any underlying ``soul'' or singular mind.\footnote{Baron Edward Thurlow is widely credited with the aphorism that “corporations have neither bodies to be punished nor souls to be condemned” (Coffee 1981).}

Similar patterns arise with respect to AI. As political theorist David Runciman observes, we have already begun ``the handover,'' delegating consequential decisions to AI systems in ways that make them function as autonomous decision-making entities within institutional life (Runciman 2023). When a bank responds to an algorithmic fraud detection system by freezing an account, when a government bases policing strategy on predictive risk scores, when millions of drivers adjust their routes based on navigation algorithms, or when hiring managers rely on automated screening to determine which candidates advance -- these are not instances of ``using a tool.'' They are instances of actors responding to another actor's acts and decisions, whether those acts or decisions are made by a corporation, a regulatory agency, a professional association, or people more generally.

These patterns suggest the following criterion: an entity functions as an actor when other actors must orient their behavior around its operations, treat its outputs as decisions or actions requiring response, and cannot easily bypass or override its influence (Zajko 2022). By this measure, certain AI systems qualify, performing roles that mirror, supplement, or even supplant the functions historically associated with established institutional actors or individuals. These sytems are experienced not as passive instruments but as entities that structure the field of action -- what sociologist Pierre Bourdieu termed the distribution of different forms of capital (economic, cultural, symbolic, informational) that determine what is possible within a given social space (Bourdieu 1986; Bourdieu 1989). AI systems now participate in this distribution by allocating attention, validating knowledge, determining access to opportunities, and structuring the behavioral incentives to which other actors must adapt. (Note that this knowledge ``validation'' is sociological, not epistemic. Even when outputs are incomplete or hallucinatory, other actors often treat them as reliable, and that reliance gives AI systems power over what counts as knowable or actionable.)

This functional definition of societal actor sets aside deeper philosophical questions about consciousness, intentionality, and moral status. We bracket these not because they are unimportant, but because they are not necessary to resolve the governance challenge at hand. Just as corporate legal personhood does not require attributing consciousness to firms, treating AI as a societal actor does not require resolving whether AI systems ``truly'' possess agency in some metaphysical or other sense. What matters for policy is that AI systems function institutionally as sources of organized decision-making that other actors cannot ignore. This is familiar in legal doctrines that treat agency and fiduciary relationships as problems of attribution, where responsibility is allocated across relational actors who exercise some degree of authority and control (DeMott 2003).

Recent work extends these attributional logics to AI, analyzing how concepts such as authority, loyalty, and third-party reliance might apply to AI agents whose choices are not reducible to ongoing human control (Riedl \& Desai 2025; Ayres \& Balkin 2025). Where such systems make consequential choices without a clear fit to traditional categories of accountability, scholars describe a growing ``responsibility gap,'' in which decisions are made but no actor fully bears responsibility (K\"onigs 2022). Recognizing AI systems as actors in this analytical sense reflects the embedding of decision-making power within system architecture itself without necessitating determinations of legal personhood or moral agency.

The criterion also clarifies useful boundaries. Markets, media ecosystems, and cultures shape behavior profoundly but lack the attributes of discrete, organized entities with bounded operations and identifiable outputs. Markets in particular coordinate behavior through decentralized interactions among many participants (actors), not through a coordinated or functionally unified system capable of producing actions that can be attributed back to a recognizable locus of agency. Markets do not ``act'' in the sociological or organizational sense. Instead, markets emerge from the aggregated behaviors of many actors and cannot initiate, execute, or be held responsible for specific decisions or rule-like outputs. In this respect, they differ from entities -- whether human, institutional, or technical -- whose patterned operations generate effects that others must treat as decisions or actions. Thus, a market is a coordination mechanism, not an actor; media is an environment, not a participant; culture is a context, not a decision-maker. AI systems, by contrast, have identifiable operational boundaries (specific algorithms, training data, deployment contexts), produce discrete outputs (classifications, recommendations, executed actions), and can be addressed, regulated, or contested as distinct entities -- even if their internal processes remain opaque.

For these reasons, clinging to a strictly anthropocentric definition of actorhood obscures more than it clarifies. It prevents us from seeing how authority is now exercised, how responsibility becomes diffused, and how the capacity to structure the field of possible actions (what Michel Foucault called \textit{pouvoir}) is embedded in AI systems rather than residing solely in human individuals or traditional institutions (Foucault 1982). Recognizing AI as an actor -- one whose capacity to act is emergent, relational, and infrastructural -- forces a more accurate account of how contemporary institutions actually operate. It makes visible the systems that increasingly govern everyday life and provides the conceptual groundwork for understanding the implications of AI's role as a fourth societal actor in Sections 3 and 4.

\subsubsection{Key Qualities Qualifying AI as an Actor}

The subset of AI systems that qualify as societal actors exhibit distinctive qualities -- properties of how modern AI is built, trained, and deployed into institutional settings. These qualities are not merely technical features; they are political assets. When an entity possesses the capability to structure human choices, distribute resources, and withhold information, it has moved beyond mere instrumentality. It is wielding power. In political theory, an entity that wields independent power over societal outcomes is, by definition, a societal actor (Peters 2016). These qualities, summarized in Table 1, help explain why other actors experience AI systems as entities whose operations must be anticipated, accommodated, or counterbalanced.

\begin{table}[h]
\centering
\caption{Societal actor qualities}
\label{tab:1}
\small
\begin{tabular}{>{\raggedright\arraybackslash}p{0.3\textwidth}>{\raggedright\arraybackslash}p{0.65\textwidth}}
\hline
\vspace{0pt}\textbf{Quality} & \textbf{Explanation} \\
\hline
\vspace{0pt}Distributed collective intelligence & Aggregates and synthesizes inputs across diverse sources and populations \\
\addlinespace
\vspace{0pt}Normative embeddedness & Encodes classifications, preferences, and value-laden assumptions that structure judgments \\
\addlinespace
\vspace{0pt}Operational opacity & Functions through processes not fully interpretable even by designers or operators \\
\addlinespace
\vspace{0pt}Infrastructural embeddedness & Occupies structurally central positions in institutional systems, becoming difficult to remove or replace \\
\hline
\end{tabular}
\end{table}

\begin{enumerate}
\item[(a)] \textbf{Distributed collective intelligence.} Contemporary AI systems aggregate and synthesize information across vast populations and data sources, producing outputs that reflect patterns no individual human could discern or replicate (Rahwan et al. 2019; Russell 2019). Large language models train on billions of documents; recommender systems learn from millions of user interactions; fraud detection systems process transactions across entire financial networks. This distributed intelligence parallels the collective intelligence exhibited by other institutional actors: electorates aggregate individual preferences into political mandates, corporations synthesize distributed knowledge into coordinated production, states compile information across bureaucracies into policy. The key similarity is that the resulting ``intelligence'' emerges from aggregation rather than individual cognition, making it irreducible to any single individual (Landemore 2013).

\item[(b)] \textbf{Normative embeddedness.} AI systems contain built-in classifications, preferences, and value-laden assumptions that structure how behaviors, people, and situations are evaluated (Noble 2018; Benjamin 2019). Training data encode historical patterns of discrimination; optimization functions privilege certain outcomes (engagement, efficiency, risk minimization) over others (equity, dignity, democratic deliberation); and classification systems impose categorical boundaries that shape institutional treatment (Obermeyer 2019). These normative commitments are rarely explicit. They emerge from technical choices about training data, objective functions, and architectural design but operate similarly to the normative frameworks embedded in legal codes, corporate cultures, or bureaucratic procedures. They structure judgment in systematic, consequential ways that persist across countless individual decisions (Jasanoff 2004).

\item[(c)] \textbf{Operational opacity.} AI decision-making processes, especially in complex machine learning models, are partially opaque and not fully controllable even to their designers or operators (Burrell 2016, Peters 2023, Holzinger 2024). Neural networks with billions of parameters develop internal representations that resist human comprehension; ensemble models combine multiple algorithms in ways that obscure which factors drove specific outcomes; systems trained on proprietary data operate as commercial secrets (Pasquale 2015). This opacity creates functional independence: once deployed, AI systems can produce actions or decisions without direct human intervention at each step, and when those decisions produce unexpected or problematic outcomes, the causal pathways often cannot be reconstructed (Vaassen 2022). The resulting gaps in oversight and accountability mean that other actors -- individuals seeking recourse, regulators attempting compliance review, competing firms trying to understand market dynamics -- must treat AI systems as entities whose behavior can be observed and responded to, but not fully predicted or controlled (Bathaee 2018).

\item[(d)] \textbf{Infrastructural embeddedness.} When AI systems become integrated into core institutional processes -- government administration, corporate workflows, critical infrastructure, financial markets -- they become structurally central in ways that make removal or radical alteration highly disruptive (Ananny \& Crawford 2018; Dargan 2024). In practice, once an AI system is embedded (for example, an algorithmic hiring system in human resources, or a routing algorithm in logistics, or an automated system in policing), organizations start to depend on it (Orlikowski 2000). Procedures are redesigned around algorithmic outputs; staff develop expertise in working with rather than around the system; and complementary technologies and processes are built assuming the AI's continued operation. This infrastructural lock-in gives AI systems organizational persistence, through which they shape not only current operations but also the design of successor systems, as institutions find it easier to iterate on existing AI architectures than to reimagine processes without them (Omidvar 2023; Hughes 1987; David 1985). In this respect, AI systems take on an enduring institutional role, rather than functioning as replaceable tools.
\end{enumerate}

\subsubsection{Key Capabilities Qualifying AI as an Actor}

While qualities describe properties of AI systems that warrant actor status, capabilities describe what AI-as-actor systems do -- the concrete functions through which they shape outcomes, allocate opportunities, and restructure institutional processes. These capabilities, summarized in Table 2, make AI not just present in social systems but consequential within them.

\begin{table}[h]
\centering
\caption{Societal actor capabilities}
\label{tab:2}
\small
\begin{tabular}{>{\raggedright\arraybackslash}p{0.3\textwidth}>{\raggedright\arraybackslash}p{0.65\textwidth}}
\hline
\vspace{0pt}\textbf{Capability} & \textbf{Explanation} \\
\hline
\vspace{0pt}Epistemic capabilities & Interpretation, classification, and curation of information at scale \\
\addlinespace
\vspace{0pt}Decisional capabilities & Structuring available choices and determining outcomes \\
\addlinespace
\vspace{0pt}Implementation capabilities & Direct execution of consequential actions without human intermediation \\
\addlinespace
\vspace{0pt}Coordination capabilities & Orchestration of behavior across distributed systems and populations \\
\addlinespace
\vspace{0pt}Norm-setting capabilities & Defining standards, structuring incentives, and shaping behavioral expectations \\
\hline
\end{tabular}
\end{table}

\begin{enumerate}
\item[(a)] \textbf{Epistemic capabilities.} Large-scale models, classifiers, and recommender systems filter, rank, and interpret information in ways that now constitute a dominant epistemic infrastructure of digital society (Pasquale 2015; Gillespie 2018). They determine which content becomes visible or credible, who is categorized as risky or trustworthy, and what patterns are deemed significant (Zuboff 2019). Search engines curate humanity's accessible knowledge; content moderation systems adjudicate what speech is permissible; credit scoring models define financial trustworthiness; diagnostic AI systems interpret medical imaging. AI systems are behind all of these, sometimes in concert. In performing these functions, AI systems occupy a role analogous to Weberian bureaucracies that create and maintain the knowledge structures through which institutions make sense of the world (Weber 1978; Porter 1995). The resulting interpretations are not passive reflections; they shape the informational horizons of billions of people and countless institutions, determining what becomes knowable, actionable, or ignorable within specific domains. This coupling of knowledge production and structural influence -- what Foucault termed the \textit{savoir-pouvoir} nexus -- positions AI systems as key sites where what counts as knowledge becomes inseparable from what becomes institutionally actionable (Foucault 1980).

\item[(b)] \textbf{Decisional capabilities.} AI systems structure the choices available to human and institutional actors, effectively functioning as decision-makers even when humans formally retain final authority (Grimmelikhuijsen \& Meijer 2022; Alon-Barkat 2023; Zeiser 2024). Recommender systems determine what users encounter, shaping consumption and attention; predictive policing tools identify suspects and allocate enforcement resources (Richardson, Schultz \& Crawford 2019; Eubanks 2018); automated hiring systems screen candidates and determine who advances; risk-scoring models shape financial and administrative decisions; treatment recommendation systems guide medical care (O'Neil 2016). These selections result from algorithmic optimization -- maximizing engagement, predicted accuracy, or operational efficiency, for example -- as a replacement for human deliberation about what constitutes a good outcome in context (Selbst 2021). They also function as decisions in the sociological sense by allocating opportunities, foreclosing alternatives, and structuring subsequent behavior by other actors who must operate within the choice architectures AI systems construct (Thaler \& Sunstein 2008). Crucially, these decisional capabilities operate at scales no human decision-maker could match, processing millions of micro-decisions that cumulatively determine what information circulates and which opportunities become accessible to whom.

\item[(c)] \textbf{Implementation capabilities.} Some AI systems can execute operational tasks autonomously, producing direct material effects without human review or intervention (Parycek, Schmid \& Novak 2024; Rizk 2025). Fraud detection systems freeze bank accounts (Khan 2025); automated moderation tools remove posts or suspend accounts; algorithmic management platforms discipline workers by modifying schedules, monitoring performance, or adjusting pay (De Stefano \& Doellgast 2023; Ajunwa, Crawford \& Schultz 2017); autonomous vehicles make real-time navigational judgments (Miller 2024); high-frequency trading systems execute financial transactions (Yacoubian 2025); automated benefits systems approve or deny social services. These actions go beyond advisory recommendations subject to human approval and are sometimes executed directly by AI systems themselves, at speeds or scales that preclude meaningful human oversight of individual cases (Howe 2025). In effect, AI systems increasingly perform a subset of what Michael Lipsky called ``street-level bureaucracy,'' enforcing rules and distributing burdens and benefits within everyday life (Lipsky 2010).

\item[(d)] \textbf{Coordination capabilities.} AI systems coordinate interactions and orchestrate behavior across complex sociotechnical systems, functioning as infrastructural hubs around which other actors must orient their strategies (Kerry et al. 2025; Algazinov, Chandra \& Laing 2025). Navigation algorithms redirect the movement of millions of vehicles across urban networks (Metz 2022); algorithmic markets influence financial flows (Yuferova 2024) and coordinate buyers and sellers, influencing price formation and resource flows (Chen 2025); ranking systems determine whose labor or creativity achieves economic visibility (Biega, Gummadi \& Weikum 2018); and workflow management systems structure how administrative tasks are sequenced and prioritized within organizations (Parker, Van Alstyne \& Choudary 2016). Once embedded at this scale, AI becomes a coordination hub -- an entity around which other actors must orient their strategies and expectations. This means AI actively shapes collective outcomes (for example, determining how drivers navigate traffic, how buyers and sellers meet, or how voters receive political news) rather than just passively executing commands (Bar-Gill \& Sunstein 2025). This coordinative role is characteristic of actors, as it involves stabilizing patterns of collective behavior and conditioning the possibilities of action across entire populations or sectors.

\item[(e)] \textbf{Norm-setting capabilities.} Through their operational logic, AI systems implicitly define and enforce norms about desirable behavior, appropriate outcomes, and legitimate categories (Yeung 2017; Noble 2018). By classifying people and objects -- scoring someone as ``low risk'' or ``high risk'', labeling content as ``relevant'' or ``spam'', ranking creators by ``quality'' or ``engagement'' -- they effectively reward certain behaviors and penalize others (Barocas \& Selbst 2016; Caliskan, Bryson \& Narayanan 2017). Users and organizations, responding to these algorithmic preferences, often adjust their behavior to align with what the system rewards (for instance, content creators alter style to fit what recommendation algorithms favor (Bucher 2018); job applicants tailor r\'esum\'es to pass automated screening; workers change habits when monitored by algorithmic management (Cotter 2019; Kellogg, Valentine \& Christin 2020)). In establishing these behavioral expectations, AI systems perform functions typically associated with regulatory bodies, professional associations, or cultural institutions: they define what is visible or invisible, credible or suspect, successful or failing within specific domains (Lessig 2006; Yeung 2018). These normative effects emerge from the accumulated behavioral adaptations that other actors undertake in response to algorithmic sorting, ranking, and optimization. The result is a form of governance -- the structuring of conduct -- that operates continuously and automatically, shaping behavior at scales that traditional institutional norm-setters could never achieve.
\end{enumerate}

Taken together, these qualities and capabilities of AI-as-actor are not simply metaphorical flourish but a way of tracking where power actually sits. They make explicit that many of the decisions that shape people's lives and experiences are now made in, through, or under the shadow of AI systems whose logic cannot be reduced to the will of any individual human.

The duck-rabbit flips: what once looked like technology -- perhaps an epochal, world-changing technology, but a mere technology nevertheless -- comes into focus as an additional pole of organized power within the basic structure of society.

Concretely, the actor-likeness of AI varies by context. In a university, algorithmic admissions, grading, or academic integrity systems can become infrastructural gatekeepers, shaping access to opportunity and institutional norms. In consumer platforms, recommender engines structure attention and public discourse through continuous ranking and curation. In welfare, finance, or insurance, claims and eligibility models can function as de facto adjudicators, triggering consequential decisions at scale with limited transparency or contestability. Even when embodied systems such as humanoid robots remain operationally narrow, their integration into service delivery can shift accountability, discretion, and control from frontline institutions to the system's designers, maintainers, and update pipeline.

However, treating AI as a societal actor does not imply that all AI systems, or even all large language models (LLMs), are functionally identical. Just as there are many state actors and many firms with different internal cultures and capabilities, there are multiple unique AI actors, shaped by architecture, training data, post-training alignment, deployment context, and deployment context. Current transformer-based LLMs may share common representational structures, but differences in fine-tuning and objective functions already produce divergent behavioral profiles; future architectures may diverge further. (Though note that some research suggests different LLMs may converge toward similar internal representations despite architectural and training differences, a claim some describe as the ``Platonic Representation Hypothesis,'' while leaving open how post-training and alignment affect outward behavior (Huh et al. 2024).) The framework developed here is agnostic to those technical specifics. It applies to any AI system, or class of systems, that satisfies the criteria for AI-as-actor set out above.

\section{A New Four Player Game: Rebalancing Power in a Four-Actor System}

\subsection{Mapping Power Among the Four Societal Actors}

A key question then is how the Four Societal Actors model -- and AI in particular -- reshapes flows of power. We understand society today as structured around a tripartite balance of powers between People, the State, and Enterprises, with each checking the others in various ways (Federalist No. 51 1788; Fukuyama 2014; Acemoglu \& Robinson 2012). In the four-actor model, AI itself checks, and is checked by, the other three societal actors, precipitating a rebalancing of powers.

Policy serves as the mechanism through which society deliberately calibrates these power relationships, establishing rules for how the actors' interactions should be structured to enable collective progress, while at times privileging a particular actor's ability to prevail in specific circumstances -- but always preventing unilateral domination or other destabilizing outcomes. A rebalancing of powers prompts fundamental reconsideration of policy choices. For example, decisions previously framed as trade-offs between state regulation and market freedom for enterprises must now account for AI's independent capacity to constrain or enable both.

\subsubsection{The Five Modalities of Power}

The levers available to each societal actor for influencing societal outcomes manifest in five power modalities: economic, informational (epistemic), informational (narrative), authoritative (consent-based), and physical (Mann 1986; Barnett \& Duvall 2005). Together, these modalities encompass how societal control is organized. They are summarized in Table 3.

The economic modality concerns allocation of resources and the incentive structures that ultimately influence how value is determined and understood (Foucault 1977; Scott 1998; Lessig 1999). Informational power splits into epistemic and narrative power: epistemic power curates what serves as fact, while narrative power exerts influence by shaping how those facts are interpreted and mobilized (Gramsci 1971; Foucault 1980; D'ancona 2017). Authoritative power exists where people accept an actor's legitimacy according to recognized governance institutions and the rule of law (Arendt 1970; Weber 1978; Habermas 1996). Physical power includes the use of tangible physical force, or the threat thereof, to influence the agency of other actors in the Hobbesian sense (Hobbes 1651/1996; Weber 1946; Arendt 1970).

Each power modality flows through various institutional ecosystems, \textit{i.e.}, the infrastructures through which power is channeled and modulated. Economic power operates through capital markets, banking systems, and supply chains; epistemic power flows through research institutions, databases, and information networks; narrative power channels through media ecosystems, social platforms, and communication systems; authoritative power manifests through democratic institutions, corporate governance structures, and legal frameworks; and physical power is exercised through military capabilities, law enforcement, and control over critical infrastructure. The design of these environments determines not just which actors can effectively exercise power, but how quickly it can be mobilized and how broadly it can reach.

\begin{table}[htbp]
\centering
\caption{The five modalities of power}
\label{tab:3}
\small
\begin{tabular}{>{\raggedright\arraybackslash}p{0.20\textwidth}>{\raggedright\arraybackslash}p{0.40\textwidth}>{\raggedright\arraybackslash}p{0.35\textwidth}}
\hline
\vspace{0pt}\textbf{Power modality} & \textbf{Description} & \textbf{Enabling ecosystems} \\
\hline
\vspace{0pt}Economic & The ability to allocate or withhold resources, capital, labor, and financial incentives & Capital markets, banking systems, supply chains \\
\addlinespace
\vspace{0pt}Informational: Epistemic & The ability to control raw data and information systems & Research institutions, databases, information networks \\
\addlinespace
\vspace{0pt}Informational: Narrative & The ability to shape knowledge, narratives, public opinion, and behavior through communication and influence & Media ecosystems, social platforms, communication systems \\
\addlinespace
\vspace{0pt}Authoritative (consent-based) & The ability to command or enforce actions based on recognized legitimacy, whether legal, institutional, or cultural & Democratic institutions, corporate governance structures, legal frameworks \\
\addlinespace
\vspace{0pt}Physical & The ability to apply or constrain movement, force, violence, and spatial access & Military capabilities, law enforcement, control over critical infrastructure \\
\hline
\end{tabular}
\end{table}

The Four Societal Actors model recognizes AI alongside the State, Enterprises, and People as concurrent power centers, each with the potential to wield power via the five power modalities. Like the proverbial 800-pound gorilla, AI does not politely ask where to sit -- it sits where it wants, and everything else in the room must adjust accordingly. In practice, the manifestation and exercise of power among and between the societal actors is relational rather than absolute, shaped by patterns of interaction rather than by intrinsic possession. Strictly speaking, this means that AI, like any of the other societal actors, might dominate all five power modalities.

At present, AI's most obvious manifestations of power are through the epistemic and narrative domains. By processing vast amounts of data and information (inclusive of incorrect information), AI can begin to govern the construction of truth and the collective understanding thereof (Zuboff 2019). Epistemic power, of course, also spills over into other modalities. Custody of data and knowledge gives economic advantage through market intelligence and other mediums, authoritative influence (agenda-setting and evidence selection in policy), and physical capability (intelligence and data feeding into targeting systems). In practice, this makes epistemic power an upstream modality that quietly amplifies downstream power across the system.

Mapping how each societal actor might manifest the power modalities enables policymakers to forecast where AI has the potential to amplify, distort, or disrupt existing power distributions (see Table 4). It also lends a practical lens to assess where intervention is needed or useful, such as correcting information asymmetries, balancing inequitable resource allocation, and preventing physical coercion.

\begin{sidewaystable}[htbp]
\centering
\caption{Manifest Power Modality Examples}
\label{tab:4}
\footnotesize  
\renewcommand{\arraystretch}{1.3}
\begin{tabular}{>{\raggedright\arraybackslash}p{0.12\textwidth}>{\raggedright\arraybackslash}p{0.12\textwidth}>{\raggedright\arraybackslash}p{0.12\textwidth}>{\raggedright\arraybackslash}p{0.12\textwidth}>{\raggedright\arraybackslash}p{0.12\textwidth}>{\raggedright\arraybackslash}p{0.22\textwidth}}
\hline
\vspace{0pt}\textbf{Power Modality / Actor} & \textbf{People} & \textbf{State} & \textbf{Enterprises} & \textbf{AI} & \textbf{Dynamic Tendency (under stress or crisis)} \\
\hline
\vspace{0pt}Economic & Labor force, consumer demand, entrepreneurship & Taxation, spending, market regulation & Capital, market control, pricing, employment & Resource flow optimization, autonomous procurement, AI agent marketplaces & Shifts toward actors with control over liquidity and infrastructure; crises accelerate concentration of capital and dependency. \\
\addlinespace
\vspace{0pt}Informational (Epistemic) & Citizen expertise, open-source knowledge, collective intelligence & Public data governance, classification systems \& statistics, research funding, education policy & Data ownership, R\&D control, proprietary algorithms, information gatekeeping & Synthetic data generation, inferential modeling, self-updating knowledge systems & Crises amplify epistemic asymmetry; AI and data-rich actors gain disproportionate authority over ``truth'' production. \\
\addlinespace
\vspace{0pt}Informational (Narrative) & Cultural norms, influencers, civil movements, arts & Public communication, media regulation, education, censorship, diplomacy & Marketing, platform design, branding, behavioral nudging & Algorithmic curation, synthetic media, virality engines, AI companions & Power shifts to actors able to weaponize attention cycles; narrative volatility spikes during emergencies and elections. \\
\addlinespace
\vspace{0pt}Authoritative (Consent-Based) & Voting, collective action, jury service & Courts, police force, national sovereignty & Contracts, organizational governance structures, leadership mandates & Delegated decision authority, reliance on AI system outputs (copilots, diagnostics) & Authority expands toward actors promising stability; overreach risks legitimacy erosion and civil pushback. \\
\addlinespace
\vspace{0pt}Physical & Mass mobilization, civil resistance \& protest, self-defense & Military, police, infrastructure control & Ownership of physical infrastructure \& industrial assets, logistics systems & Robotics, digital twins with real-world control, cyber-physical control & Crises trigger coercive consolidation; states regain dominance but risk societal fragmentation or automation backlash. \\
\hline
\end{tabular}
\end{sidewaystable}

\subsubsection{The Role of Legitimacy for AI-as-Actor}

Understanding AI's distinctive source of legitimacy is crucial for policymaking because it unveils the fragility of its own power, as well as the potential instability it introduces into the Four Societal Actors ecosystem. AI's legitimacy is reinforced by structural incentives, such as efficiency pressures, resource scarcity, and political demands for ``neutral'' decisions. Delegation to AI systems is therefore encouraged, even if the automation bias and opacity that accompany such delegation mean that legitimacy is often procedural rather than substantive.

The other three societal actors, by contrast, stand on socially validated foundations. The State's legitimacy flows from constitutional order and rule of law; Enterprises earn legitimacy through market competition, contractual trust, and shareholder accountability; People derive legitimacy through collective action and democratic participation such as protests, voting, and organized movements (Tocqueville 1840/2000). These legitimacy sources are institutionally embedded, with established mechanisms for renewal, contestation, and repair when eroded (Weber 1978; Habermas 1984).

But unlike voters who elect governments or consumers who validate companies through purchases, AI gains legitimacy through a different path: we delegate our judgment to it. When we adopt AI systems for consequential decisions -- credit scoring, hiring, medical diagnosis -- we implicitly transfer legitimacy to these systems, accepting their outputs as objective, rational, and efficient (Habermas 1996; Pasquale 2015; Sunstein 2025). This represents an extension of Weber's concept of legal-rational authority, but without the institutional accountability mechanisms Weber's framework assumes (Weber 1978). AI's legitimacy draws authority from design confidence, apparent neutrality, and demonstrated efficiency -- but these are brittle foundations. A single high-profile algorithmic failure can cascade into systemic trust collapse in ways that traditional institutional actors can often survive through reform or leadership change. Consider the following examples:

\begin{enumerate}
\item[(a)] \textbf{UK Welfare Fraud Algorithm (2024).} The UK Department for Work and Pensions' (DWP) automated fraud-detection system wrongly flagged around 200,000 legitimate benefit claims as potential fraud. The model disproportionately targeted vulnerable households and offered no meaningful route for claimants to contest the decision because frontline staff themselves relied on the algorithm's opaque outputs. Public criticism forced the DWP to pause parts of the system, but the underlying logic remains embedded in welfare administration (Walker 2024). This case shows how delegating administrative authority to AI can erode institutional legitimacy: when an automated system fails, neither the outputs nor the harm can be repaired through ordinary bureaucratic or legal mechanisms. The State's legitimacy is strained, not as a result of intentional overreach, but because the institutional chain of accountability collapses into code.

\item[(b)] \textbf{Amazon's Hiring Algorithm (2018).} Amazon discovered its AI recruiting tool systematically downgraded resumes containing the word ``women's'' (as in ``women's chess club captain'') because it learned from historical hiring patterns in a male-dominated industry. The company disbanded the tool, but not before it had influenced real hiring decisions. The legitimacy Amazon had delegated to this system, that is, trusting it to screen resumes ``objectively,'' evaporated instantly, with no pathway to restoration beyond scrapping it entirely (Dastin 2018). This case illustrates how AI's legitimacy, unlike institutional legitimacy, cannot be reformed from within; it must be reconstructed externally.

\item[(c)] \textbf{UK A-Level Grading Algorithm (2020).} During COVID-19, the UK government used an algorithm to assign grades to students whose exams were cancelled. The system systematically downgraded students from disadvantaged schools based on historical performance patterns. Public outcry forced the government to abandon the algorithm entirely and revert to teacher assessments. Here, the State attempted to delegate its educational authority to an AI system, but the algorithm's opacity made it impossible to defend the decisions through normal administrative or legal processes. The only solution was complete reversal (Hao 2020). This demonstrates how algorithmic legitimacy, when it fails, can undermine the legitimacy of the delegating institution itself.

\item[(d)] \textbf{Healthcare AI Bias (2019).} A widely-used healthcare algorithm was found to systematically underestimate the medical needs of Black patients because it used healthcare spending as a proxy for health needs, and Black patients historically had less money spent on their care due to systemic inequalities. Hospitals had delegated critical resource allocation decisions to this system, trusting its ``objective'' risk calculations. Unlike a discriminatory hospital administrator who could be fired or retrained, the algorithm required fundamental redesign of its objective function (Obermeyer et al. 2019). This case reveals how AI systems can embed and amplify existing inequalities while appearing neutral, creating what Habermas would call a ``legitimation crisis'' without communicative mechanisms for resolution (Habermas 1975).
\end{enumerate}

These cases reveal critical asymmetries. When traditional actors fail, they can explain, apologize, reform, and rebuild trust through communicative processes -- the State through public hearings and policy changes; Enterprises through leadership accountability and market correction; and People through new organizing strategies. AI systems cannot do this. They have no voice to explain, no capacity to promise reform, and no mechanism for rebuilding trust except through external human intervention. Table 5 summarizes each societal actor's source of legitimacy and fragility factors.

Hence, the limitations of AI's ``delegated procedural legitimacy.'' Without such mechanisms for remedy, AI's delegated legitimacy remains dangerously fragile. Consent is procedural rather than deliberative, automated rather than reasoned. We delegate our judgment through default acceptance, becoming functionally obedient to AI through automation bias and the convenience of delegating cognitively demanding tasks (Parasuraman \& Manzey 2010; Kahneman 2011).

\begin{table}[htbp]
\centering
\caption{Legitimacy: Sources and fragility factors}
\label{tab:5}
\small
\begin{tabular}{>{\raggedright\arraybackslash}p{0.12\textwidth}>{\raggedright\arraybackslash}p{0.25\textwidth}>{\raggedright\arraybackslash}p{0.25\textwidth}>{\raggedright\arraybackslash}p{0.30\textwidth}}
\hline
\vspace{0pt}\textbf{Actor} & \textbf{Source of Legitimacy} & \textbf{Example} & \textbf{Fragility Factor} \\
\hline
\vspace{0pt}People & Democratic consent, civic participation, collective action & Elections, protests, civil deliberation, social movements & Apathy, misinformation, political polarization eroding collective agency \\
\addlinespace
\vspace{0pt}State & Constitutional order, legal-rational authority, rule of law & Judicial review, bureaucratic process, electoral accountability & Corruption, authoritarianism, or legitimacy crisis undermining public trust \\
\addlinespace
\vspace{0pt}Enterprises & Market competition, contractual trust, fiduciary duty & Shareholder accountability, market discipline, legal enforcement, reputation & Scandals, monopolistic behavior, fraud eroding market or consumer trust \\
\addlinespace
\vspace{0pt}AI & Delegated procedural legitimacy (perceived objectivity, efficiency, and neutrality) & Performance metrics, audit mechanisms (where they exist), user acceptance & Automation bias, opacity, algorithmic failures, lack of communicative validation or moral agency \\
\hline
\end{tabular}
\end{table}

AI's legitimacy fragility has profound policy implications. Conventional regulatory approaches presuppose actors whose legitimacy is self-sustaining and can be constrained or redirected through familiar legal tools. By contrast, AI's legitimacy is entirely derivative, borrowed from the humans and institutions that design, deploy, and rely on it. When that chain of delegation breaks, there is no inherent repair mechanism through which the system can explain itself, acknowledge error, or seek renewed acceptance -- no ``communicative'' path back to trustworthiness (Habermas 1984). Thus, in Habermasian terms, AI embodies a new form of non-communicative, procedural legitimacy, through which systems derive authority from formally correct procedures and delegated mandates, even though they cannot themselves participate in the communicative practices that ground legitimacy.

To remain viable, this procedural legitimacy must be embedded in communicative infrastructures -- oversight bodies, contestation channels, and auditing regimes -- that reconnect algorithmic outputs to human justification. AI governance must therefore prioritize institutional arrangements for ongoing legitimacy validation that makes AI's delegated authority subject to the same communicative validation that legitimate authority requires, so that decisions rendered in code remain answerable to the same justificatory standards that ground legitimate human authority, even as decision-making shifts to algorithmic systems (Rahwan 2018; Raji et al. 2020).

Unlike traditional software, AI systems do not merely execute predefined instructions. They generate classifications, predictions, and recommendations that function as substitutes for human judgment. The authority we delegate to AI is thus qualitatively different; we are outsourcing discretion, not just processing. Delegated procedural legitimacy therefore behaves differently from traditional forms of technological delegation. When a database fails, we repair the tool while the underlying institutional authority remains intact. When an algorithmic decision-maker fails, the failure directly compromises the legitimacy of the decision itself, precisely because the system has functioned as the decision-maker.

This dynamic makes legitimacy analysis central to the Four Societal Actors model. Recognizing AI as an actor whose authority is derivative reveals how its failures create governance gaps that neither the State, Enterprises, nor People can readily fill. Treating AI as a power node within the model highlights this distinctive risk: its authority is fragile yet rapidly scalable, and it cannot be stabilized through the communicative or institutional practices that renew the legitimacy of the other actors.

Framing AI in this way also clarifies why the familiar regulation-versus-innovation binary is so unsatisfying. It obscures how AI's unusual legitimacy structure generates policy demands that do not map onto conventional technology governance. Once AI is understood as an actor with its own legitimacy dynamics, the central task for policymakers shifts from deciding whether to regulate AI's capabilities to designing institutions capable of incorporating powerful digital actors without eroding the legitimacy of existing institutions or allowing AI's authority to expand beyond accountability.

That said, reframing AI as a societal actor does not, by itself, resolve questions of accountability when AI-mediated decisions cause harm. It does, however, clarify the institutional design space in which that question must be answered. It also allows for distinguishing, for example, between purely human-centered liability regimes (vicarious or strict liability for deployers) and hybrid arrangements in which AI systems are treated as loci of agency that trigger no-fault compensation schemes, as some scholars have suggested in debates over autonomous-vehicle accidents and insurance-based models of responsibility (Blackman 2025). Developing concrete accountability architectures for AI actors is therefore a central agenda for future work, one that this article does not attempt to settle but seeks to make analytically tractable.

\subsection{Policy and Governance Implications of the Four Societal Actors Model}

The Four Societal Actors model unveils AI governance as fundamentally a coordination problem among actors with competing interests and asymmetric capabilities (Ostrom 1990; Jasanoff 2016). Rather than viewing policy as a set of rules imposed on technology, we can understand it as moves in an ongoing game, where each actor attempts to maintain or expand their power across the modalities, while other actors deploy checks to prevent domination.

This game-theoretic lens serves two purposes. First, it helps diagnose where power imbalances emerge as AI capabilities advance, revealing which actor-modality combinations might become dangerously concentrated or vulnerable to exploitation. Second, it provides a framework for designing and evaluating policy interventions not as isolated regulations, but as strategic moves that shift power dynamics across the entire system (Schelling 1960; Dafoe 2018).

\subsubsection{Using Power Dynamics as a Policy Diagnostic}

Power, of course, is not static; hence, the phrase ``power dynamics.'' Which actor is able and chooses to exploit a particular power modality at any given time ebbs and flows. Crises, technological advances, and environmental shocks can modulate a given power modality's impact, as well as a given actor's ability to exploit it (Pierson 2000; Klein 2007). For example, liquidity crises might centralize economic power and data asymmetries amplify epistemic power, and fake news might distort AI's narrative power (Benkler, Faris \& Roberts 2018). For instance, the 2008 financial crisis centralized economic power in institutions deemed ``too big to fail'' (Tooze 2018), while the COVID-19 pandemic shifted epistemic power toward public health agencies and pharmaceutical companies with access to clinical data (Jasanoff 2021). The 2016 U.S. election demonstrated how narrative power could be weaponized through social media platforms, whereas the Cambridge Analytica scandal revealed how data asymmetries translate into political influence (Wardle \& Derakhshan 2017).

Power modalities are also not static categories. In this framework, power modalities are strategic levers that actors deploy in response to one another. Our paradigm aims to make power shifts between players feel observable and modelable, bringing the discussion of power out of the abstract.

Understanding these new four-way power dynamics is critical to designing AI policy that is effective and robust to AI's swift evolution and growing impact. Mapping how each societal actor might check the power of the other societal actors surfaces potential imbalances introduced by AI, as well as power interdependencies, amplification, or cancellation between actors (Lukes 2005). This sets the groundwork for exploring effective policy interventions and tracking the overall balance of power between actors. See Table 6 for examples of power checks between actor pairings.

\begin{sidewaystable}[htbp]
\centering
\caption{Directional Power Checks Examples}
\label{tab:6}
\small
\renewcommand{\arraystretch}{1.3}
\begin{tabular}{>{\raggedright\arraybackslash\bfseries}p{0.12\textwidth}>{\raggedright\arraybackslash}p{0.20\textwidth}>{\raggedright\arraybackslash}p{0.20\textwidth}>{\raggedright\arraybackslash}p{0.20\textwidth}>{\raggedright\arraybackslash}p{0.20\textwidth}}
\hline
\vspace{0pt}\textbf{From \textbackslash{} To} & \textbf{People} & \textbf{State} & \textbf{Enterprises} & \textbf{AI} \\
\hline
\vspace{0pt}People & -- & Voting, protest, civil disobedience, jury nullification, public opinion & Consumer choice, labor strikes, reputational pressure, litigation & Refusal to adopt, prompt steering, adversarial use, sectoral/ unionized/ trade-specific ``war'' against AI use \\
\addlinespace
\vspace{0pt}State & Law enforcement, surveillance, coercion & -- & Regulation, antitrust enforcement, taxation, procurement control & Legislation, bans, export controls, nationalization, kill switches \\
\addlinespace
\vspace{0pt}Enterprises & Product design, pricing, availability, targeted marketing & Lobbying, regulatory capture, privatized infrastructure (e.g., cloud), political donations & -- & AI development, ownership, training data control, capability throttling \\
\addlinespace
\vspace{0pt}AI & Fostering ``monoculture'', manipulating choices, eroding attention, polarizing norms & Speed outpacing regulation, undermining legitimacy via misinformation or systems gaming & Automating decisions, price manipulation, labor displacement, market disruption & -- \\
\hline
\end{tabular}
\end{sidewaystable}

\subsubsection{From Diagnosis to Intervention: Designing Policy Levers}

To demonstrate how these dynamics reveal policy gaps and inform interventions, consider three illustrative power shifts informed by recent events.

\begin{enumerate}
\item[(a)] \textbf{Epistemic and Narrative Power: AI vs. State.} AI's recommender systems and synthetic content generation shape public understanding faster than state regulatory or communicative capacity can respond. The balance of power moves sharply in favor of AI because its speed and scale in processing and disseminating information functionally outcompetes deliberative democratic processes (Rosa 2013; Zuboff 2019). This reflects a broader shift in epistemic authority. Once information processing exceeds human deliberative bandwidth, public institutions risk becoming reactive rather than taking leadership in agenda setting (Floridi 2014).

When AI systems curate what citizens see and believe before public institutions can even identify emerging narratives, the State's authoritative power becomes downstream of AI's informational power. This undermines ``communicative legitimacy'' -- the process through which public deliberation confers legitimacy on collective decisions (Habermas 1984). Policy interventions might therefore reinforce opportunities for public deliberation through measures like: transparency requirements for algorithmic curation, friction mechanisms that slow viral spread to allow fact-checking, public media infrastructure that operates outside algorithmic gatekeepers, and appointed councils that can rapidly assess and respond to AI-driven information distortions.

\item[(b)] \textbf{Authoritative Power: State vs. People.} During crises -- pandemics, security threats, economic shocks -- States historically expand authoritative power through emergency measures that bypass normal democratic checks. AI amplifies this dynamic by enabling surveillance, prediction, and control at unprecedented scale and speed. The balance of power shifts in favor of the State, creating temporary dominance that can calcify into permanent authority absent post-crisis reversion mechanisms (Scheuerman 2006; Feldstein 2019).

When emergency AI deployments (contact tracing apps, predictive policing expansions, automated benefit denials, and so on) persist beyond crises, they normalize exceptional authority as routine governance. Authority divorced from ongoing consent becomes indistinguishable from coercion, eroding the legitimacy foundation it requires (Arendt 1970; Peterson 1999). Policy must therefore embed legitimacy safeguards within the power modality itself, for example: sunset clauses that automatically terminate emergency AI systems, independent review boards with power to revoke authorizations, mandatory public reporting on AI-enabled state actions, and clear legal frameworks distinguishing temporary crisis response from permanent capability expansion.

\item[(c)] \textbf{Economic Power: Enterprise vs. People.} AI-enabled platforms algorithmically restructure labor markets, from gig work pricing to hiring decisions to performance monitoring, faster than workers can organize collective responses. The balance of powers moves toward Enterprise as algorithmic management fragments and individuates what were once collective bargaining positions. Algorithmically mediated workplaces tighten managerial control while weakening worker contestation mechanisms (Kellogg, Valentine \& Christin 2020). People's agency reduces to consumer choice rather than co-production of economic arrangements (Rahman \& Thelen 2019).

Technological change does not automatically benefit labor, and power asymmetries tend to determine who captures productivity gains (Acemoglu \& Johnson 2023). When AI systems optimize Enterprise objectives like efficiency, cost reduction, and predictability without mechanisms for worker input or benefit-sharing, economic power concentrates even as aggregate productivity rises. Policy interventions should therefore recognize and address resource flow asymmetries, for example: portable benefits not tied to single employers, algorithmic transparency in wage-setting and scheduling, worker access to performance data, collective bargaining rights that extend to algorithmic management systems, and potentially new organizational forms (platform cooperatives, stakeholder corporations) that distribute AI's economic benefits more broadly.
\end{enumerate}

We present a more generalized depiction of how power dynamics might unfold and inform both reactive and proactive policy design in Table 7. The idea is that, at any given moment, two societal actors hold a particular power modality in tension between them, like opposing forces in dynamic equilibrium.\footnote{Or like opposing beams locked in contest, which, as fans of a certain young wizard might recognize, replicate a kind of institutional \textit{priori incantatem}, where opposing forces with common origins lock into direct contest, neither able to disengage until one yields or external intervention breaks the deadlock.} Under stable conditions, one actor typically maintains an advantage in wielding that modality's power, but appropriate checks and balances prevent unchecked dominance, at least in theory. Exogenous events or technological developments can alter the balance, as indicated by direction power shifts. Policy interventions then function as additional external forces that can strengthen one actor's hold, constrain the other's reach, or fundamentally reconfigure the contest, sometimes determining whether shifts become permanent new equilibria or temporary perturbations. Although not reflected in the table, volatility scores could also be assigned to indicate how rapidly or violently any given power in balance might destabilize, with higher scores suggesting greater urgency for policy attention.

\begin{sidewaystable}[htbp]
\centering
\caption{Examples of power dynamics and their policy implications}
\label{tab:7}
\footnotesize
\renewcommand{\arraystretch}{1.2}
\begin{tabular}{>{\raggedright\arraybackslash}p{0.13\textwidth}>{\raggedright\arraybackslash}p{0.11\textwidth}>{\raggedright\arraybackslash}p{0.10\textwidth}>{\raggedright\arraybackslash}p{0.18\textwidth}>{\raggedright\arraybackslash}p{0.10\textwidth}>{\raggedright\arraybackslash}p{0.26\textwidth}}
\hline
\vspace{0pt}\textbf{Selected Conflict Pair} & \textbf{Power Modality in Tension} & \textbf{Stable Conditions: Dominant Actor} & \textbf{Societal Event or Shock Conditions} & \textbf{Directional Power Shift} & \textbf{Potential Policy Interventions to Restore Balance} \\
\hline
\vspace{0pt}Enterprise $\leftrightarrow$ People & Economic & Enterprise & Market shocks and automation crises concentrate resource control in corporate actors. & + toward Enterprise & Call for redistributive policy and labor safeguards. \\
\addlinespace
\vspace{0pt}AI $\leftrightarrow$ State & Epistemic & AI & AI systems outperform bureaucratic capacity for data analysis. & + toward AI & Establish epistemic oversight councils to prevent information monopoly. \\
\addlinespace
\vspace{0pt}AI $\leftrightarrow$ People & Narrative & AI & Algorithmic virality and synthetic media outpace civic narrative formation. & + toward AI & Strengthen public media literacy and disclosure norms. \\
\addlinespace
\vspace{0pt}State $\leftrightarrow$ People & Authoritative & State & Emergency powers expand during crises. & + toward State & Integrate sunset clauses and independent review to prevent normalization of exceptional authority. \\
\addlinespace
\vspace{0pt}State $\leftrightarrow$ Enterprise & Physical & State & Infrastructure and coercive capacity remain state-dominant. & + toward State & Partnership frameworks should clarify boundaries on surveillance and cyber enforcement. \\
\hline
\end{tabular}
\end{sidewaystable}

These power shifts may be desirable or not; they may be temporary or enduring. Depending on societal preferences and policy goals, interventions may seek either to restore the previous status quo or deliberately reset it to a new balance. This framework thus serves as both a diagnostic tool (revealing where power is concentrating or fragmenting) and a design tool (suggesting which policy levers might restore or reconfigure balance).

Collectively, the examples presented above highlight a broader principle: when it comes to AI, effective policy intervention requires understanding which powers are shifting out of balance, why, and which countervailing forces can restore or reset the balance. The practical risk is mistaking temporary power surges for structural change; without preset decision-making apparatus or diagnostics, policymakers intervene either too early or too late. Understanding AI governance as an ongoing game among four actors also underscores why heavy-handed AI regulation is likely to fail; its propensity for lacking adaptability and becoming stale is high.

Successful AI policy frameworks must therefore embed dynamic monitoring and adaptive capacity. This means establishing systems for continuous monitoring of power distribution across modalities, such as tracking which actor-modality combinations are concentrating or fragmenting; assessing whether shifts are gradual and adaptive or rapid and destabilizing; mapping how changes in one modality cascade to others (for instance, epistemic power concentration can enable narrative control, which in turn erodes authoritative legitimacy); and anticipating second-order effects where well-intentioned interventions create new imbalances or unintended consequences elsewhere in the system.

Monitoring alone is insufficient without trigger conditions that signal when intervention is needed. These triggers might include quantitative thresholds (market concentration indices, epistemic power asymmetries exceeding defined levels, volatility scores spiking above baseline) or qualitative indicators (systematic failure of directional checks between actors, erosion of social capital, collapse of legitimacy). Policy frameworks must also include feedback loops where affected actors can signal when checks are failing and demand rebalancing to ensure that power imbalances are detected not just through abstract metrics but through the lived experience of those subject to them.

Finally, frameworks require adaptive capacity to reallocate functions between centralized and distributed governance as conditions change. When AI capabilities advance rapidly and create systemic risks, temporary centralization may be needed to prevent fragmentation and race-to-bottom dynamics. When governance becomes overly rigid and stifles beneficial innovation, authority must devolve to enable experimentation and competition. This adaptive capacity depends on maintaining a diverse portfolio of policy levers that can strengthen weakened actors or constrain overconcentrated ones depending on the diagnosis.

While a comprehensive taxonomy of policy levers is beyond the scope of this article, existing frameworks provide useful starting points. One such foundation categorizes interventions along dimensions including economic instruments (taxes, subsidies, tradable permits), command-and-control regulations (standards, bans, mandates), information-based tools (disclosure requirements, labeling, audits), rights-based mechanisms (liability rules, property rights, procedural rights), and institutional interventions (oversight bodies, certification schemes, participatory governance) (P\'erez \& Xie 2018; MacGregor \& Madsen 2018). Applied to the Four Societal Actors framework, these levers can be mapped to their effects across power modalities, for example: information tools primarily affect epistemic and narrative power; economic instruments reshape resource flows; rights-based mechanisms strengthen People's capacity to check other actors; institutional interventions build State capacity while potentially enabling Enterprise or AI actors to capture regulatory processes.

The key insight is that no single lever rebalances power alone, and attempting to engineer perfect equilibrium through comprehensive ex ante regulation risks brittle failure. Here, the game metaphor is apt: like any complex system with multiple actors, AI governance resists top-down control (Gall 2002). Interventions in one area produce compensating adjustments elsewhere, likely in unexpected ways, as Enterprises route around regulations, People find workarounds, States expand authority in adjacent domains, and AI systems adapt to constraints. Effective AI policy therefore requires coordinated interventions across multiple modalities simultaneously, with continuous feedback to detect whether desired rebalancing is occurring or whether the system is adapting around the intervention. For instance, addressing AI's epistemic dominance necessitates not just transparency requirements (information lever) but also public AI infrastructure (market structure lever), epistemic councils (institutional capacity lever), and individual contestability rights (rights-based lever) -- each reinforcing the others while preventing any single point of failure.

Such a dynamic approach prevents the brittleness that comes from assuming power distributions remain static. It acknowledges that AI governance is not a problem to be solved once, but an ongoing institutional challenge requiring vigilance, adaptation, and capability to rebalance power as circumstances evolve. In the next section, we introduce evaluative criteria for determining whether power rebalancing is moving society toward sustainable equilibrium, where both order and opportunity are preserved, or toward dangerous extremes of concentrated control or chaotic fragmentation.

\section{Winning the Game: Institutional Design for AI Governance}

If the conflict dynamics outlined in the previous sections describe how the game of power is played among the four societal actors, the next question is what it would mean for AI governance to ``win.'' Currently, the AI policy discourse is trapped in binary deadlock between innovation and regulation, acceleration and safety. Viewed through the lens of political theory, however, this is not a novel modern dilemma, but a recurrence of a foundational tension in constitutional design between liberty and power.

\subsection{The Historical Mirror: Jefferson, Hamilton, and the AI Debate}

The friction between those seeking order and those seeking unchecked development mirrors the early American debates on federalism. Thomas Jefferson argued for distributed sovereignty, prioritizing the preservation of liberty and the ability of local communities to adapt (Jefferson 1785). In the context of today's AI policy debates, this aligns with the position of modern techno-optimists, who argue that freedom to operate is essential to maximizing human flourishing (Andreessen Horowitz 2023).

Conversely, Alexander Hamilton advocated for centralized authority, arguing that without a cohesive federal power to manage currency, defense, and commerce, the system would fracture (Hamilton 1788). This parallels the warnings of the so-called ``Godfathers of AI'', who argue that without strong, centralized regulatory guardrails, AI itself introduces catastrophic risks that threaten the system's stability (Bengio et al. 2024).

The fundamental mistake in this diametric framing is viewing these positions as mutually exclusive. The brilliance of the American experiment was that, instead of choosing one view to prevail, the founders refused to choose at all. In constitutional terms, they embedded both impulses into the same design through separation of powers, federalism, and rights-based constraints so that liberty and power would check and sustain one another over time (Marbury v. Madison 1803). As scholars of American political development have noted, many of the most durable checks on power then emerged not from a single master plan but through incremental, path-dependent evolution of institutions and practices that, often unintentionally, further insulated individuals from potential tyranny (Skowronek 1982; Woodard 2016). Unlike the incoherent regulatory patchwork described in Section 2, this institutional patchwork was anchored in a shared constitutional project of checking power and preserving liberty.

\subsection{A Madisonian Architecture for AI}

The solution to the Jefferson-Hamilton tension was not a compromise, but an architecture. James Madison led the design of a system that institutionalized a productive tension between liberty and power. He recognized that pure centralization creates brittle authoritarianism, while pure distribution fragments into chaos (Federalist No. 10 1788; Federalist No. 51 1788; Weiser 2001; Balkin 2016).

To apply this to AI policy, it is useful to translate these constitutional concepts of liberty and power into more tangible system dynamics. Precisely because contemporary AI systems now perform functions once reserved to human institutions -- analyzing, deciding, coordinating at scale -- the distinctively human task shifts toward designing the constitutional environment in which those capabilities operate. Liberty, in this context, maps to \textbf{aspiration} -- the capacity to evolve, innovate, and flourish. Power maps to \textbf{stability} -- the capacity to maintain systemic coherence and prevent existential ruin.

At this constitutional or meta-governance layer, the emergence of AI as a durable actor also blurs the familiar line between ``public'' and ``private'' governance. The more informative question becomes how authority, rights, and contestability are allocated across domains and societal functions -- health, education, mobility, housing, energy, finance, security -- rather than along legacy categories such as commercial activity versus public good. The design task therefore becomes to structure those domains so that no single actor can become the final arbiter everywhere at once.

If Madison were alive today -- and perhaps an AI enthusiast -- he might propose a reframing of the regulation-versus-innovation stalemate as a federated governance design challenge, where a mosaic of policy mechanisms allocate power and provide for checks and balances based on where they best serve the societal system's overall health, shifting the focus from regulating AI systems as technology to managing power relationships among AI, People, States, and Enterprises. A Madisonian architecture for AI governance requires two complementary governance layers that foster productive tension:

\begin{enumerate}
\item[(a)] \textbf{Distributed governance: engines of aspiration.} Distributed systems fuel innovation and allow people, individually, and collectively, to aspire. Aspiration emerges from distributed functions that enable experimentation, competition, and local autonomy. Just as federalism prevents the State from capturing all spheres of life (New York v. United States 1992; Printz v. United States 1997), and antitrust law prevents Enterprises from monopolizing markets (Standard Oil Co. of N.J. v. United States 1911; Sherman Antitrust Act 2018), distributed mechanisms prevent AI from becoming the sole locus of decision-making authority within any given domain (Sabel \& Zeitlin 2008). A few examples of potential distributed governance mechanisms at the operational level include:

\begin{enumerate}
\item[(i)] \textbf{Algorithmic impact assessments with local veto power.} Analogous to environmental impact statements that allow communities to reject State projects, giving People the authority to refuse AI deployment in their jurisdictions. This combination can be understood as an expression of algorithmic sovereignty: the capacity of democratic communities to determine whether and under what conditions algorithmic systems may be deployed within their jurisdictions, rather than treating such deployment as a purely technocratic or market-driven inevitability (Reviglio \& Agosti 2020).

\item[(ii)] \textbf{Competing AI development pathways (open-source, public, private models).} Mirrors how market competition checks Enterprise power. This analogizes to the New Deal-era understanding of antitrust as a tool for disciplining concentrated private power, not merely for maximizing allocative efficiency. Historians and contemporaneous officials describe antitrust policy under Franklin D. Roosevelt as part of a broader project of ``guided competition,'' aimed at preserving pluralism in economic organization and preventing private actors from acquiring quasi-sovereign control over essential infrastructure (Roosevelt 1938; Leuchtenburg 1963). Contemporary scholarship has drawn on this tradition to argue for maintaining institutional diversity in digital and platform governance, where competition serves not only efficiency goals but also constitutional and democratic values (Wu 2018).

\item[(iii)] \textbf{Mandatory human override provisions in high-stakes domains.} Similar to corporate boards' ability to overrule executives.

\item[(iv)] \textbf{Sunset clauses on AI decision-making authority.} Requiring periodic reauthorization, just as emergency State powers expire unless renewed.
\end{enumerate}

\item[(b)] \textbf{Centralized governance: anchors of stability.} Distributed systems handle systemic shocks poorly. Stability primarily emerges from centralized functions that prevent destabilizing power concentrations and catastrophic risks. Just as constitutional limits constrain State power and corporate law imposes fiduciary duties on Enterprises, centralized mechanisms impose systemic constraints on AI actors across domains (Helbing 2013). A few examples of potential centralized governance mechanisms to consider include:

\begin{enumerate}
\item[(i)] \textbf{Fiduciary obligations for AI deployers.} Analogous to corporate directors' duties to shareholders (Easterbrook \& Fischel 1991; Miller 2013), requiring those who deploy AI actors to act in the interest of affected People, not just efficiency. Similar approaches have arisen in the data governance context. For example, India's Digital Personal Data Protection Act of 2023 defines ``data fiduciary'' as the entity that determines the purposes and means of processing personal data, and some have argued that digital platforms and algorithmic intermediaries should be treated as fiduciaries toward their users (Balkin 2016).

\item[(ii)] \textbf{Independent AI oversight bodies with subpoena power.} Mirrors judicial review of State actions or inspector generals, creating institutional capacity to audit and sanction AI actors (Mashaw 1985; Sunstein 1990). Comparable logics have informed some contemporary AI governance proposals that emphasize the need for independent oversight bodies capable of auditing algorithmic systems, compelling disclosure, and imposing sanctions where AI-mediated decisions produce unlawful or harmful outcomes (Yeung 2018).

\item[(iii)] \textbf{Prohibited decision domains (constitutional-style constraints).} Just as States cannot infringe certain rights and Enterprises cannot engage in certain monopolistic practices, AI actors could be categorically barred from domains requiring irreducibly human judgment (e.g., criminal sentencing, asylum decisions).

\item[(iv)] \textbf{Public AI infrastructure.} Analogous to public utilities or state capacity-building, ensuring States and People possess independent AI capabilities rather than depending entirely on Enterprise-controlled AI actors.
\end{enumerate}
\end{enumerate}

The key difference from traditional technology governance is that these mechanisms check institutional power, first and foremost. Of course, over-distribution can fragment opportunities and create race-to-bottom dynamics that ultimately constrain aspiration (Vogel 1995). On the other hand, over-centralization can harm stability by creating brittle single points of failure or legitimacy crises when central authority makes errors (Scott 1998; Taleb 2012). The federalist architecture makes both dimensions explicit and treats them as equally legitimate policy objectives, enabling systematic rather than ideological trade-off evaluation (Sen 1999). This echoes constitutional design choices: separation of powers and federalism deliberately divide authority to prevent tyranny, while bills of rights constrain both centralized and decentralized actors from abusing that authority.

\subsection{The Scoreboard: The Aspiration-Stability Matrix}

With federalism as the architecture, it is possible to assess efficacy via two measurable dimensions:

\begin{enumerate}
\item[(a)] \textbf{Aspiration.} The key question being asked is: Is the system worth holding together? Aspiration reflects society's need for progress, opportunity, and space for individuals and communities to pursue meaningful aims. It represents the manifestation of preserved liberty. Without aspiration, what follows is stagnation, lost human potential, and the erosion of social legitimacy that comes when systems serve only their own perpetuation. This echoes John Locke's foundational claim that government legitimacy derives from protecting both security (preserving life and property) and liberty (enabling pursuit of individual and collective aims) (Locke 1690/1988). Thus, aspiration requires conditions for human flourishing, not just in aggregate but distributively, meaning: people and communities can act and make choices without undue constraint (freedom to act); benefits and resources flow broadly rather than concentrating (opportunity distribution); the environment encourages constructive, diverse, future-oriented solutions (innovation climate); and trust between actors enables collaboration toward shared goals (social capital resilience).

\item[(b)] \textbf{Stability.} The key question being asked is: Can the system hold together? Stability is easily mistaken for safety or stagnation. However, in this framework, it represents the manifestation of balanced power between dynamic actors. This means: power is not overly concentrated in one societal actor but shared (order without domination) (Pettit 1997); feedback loops between actors are intact and not eroding (directional checks); and the system can absorb shocks without collapsing (low volatility).
\end{enumerate}

When graphed against each other, these dimensions yield four quadrants, which serve as our AI policy ``scoreboard.'' See Table 8.

\begin{table}[ht]
\centering
\caption{Aspiration-Stability Matrix}
\label{tab:aspiration-stability}
\small
\begin{tabular}{p{0.45\textwidth}|p{0.45\textwidth}}
\multicolumn{2}{c}{} \\[0.5em]
\multicolumn{2}{c}{\textbf{High Stability ↑}} \\[0.3em]
\vspace{0pt}\textbf{Quadrant I: Surviving not Thriving}
\begin{itemize}[leftmargin=*, nosep, topsep=0pt]
\item Ordered \& predictable, but power concentrated
\item Over-regulated or authoritarian environment
\item Limited innovation/opportunity
\item Goal: open space for aspiration
\end{itemize}
&
\vspace{0pt}\textbf{Quadrant II: High Potential}
\begin{itemize}[leftmargin=*, nosep, topsep=0pt]
\item Broad opportunity \& balanced power
\item Reap benefits of AI, critical safeguards in place
\item Desirable condition in most scenarios
\item Goal: preserve and adapt
\end{itemize}
\\[1em]
\hline
\vspace{0pt}\textbf{Quadrant IV: Fragile Collapse}
\begin{itemize}[leftmargin=*, nosep, topsep=0pt]
\item Severe power imbalances or shocks
\item Low trust among societal actors
\item Low opportunity
\item Goal: emergency stabilization
\end{itemize}
&
\vspace{0pt}\textbf{Quadrant III: Creative Chaos}
\begin{itemize}[leftmargin=*, nosep, topsep=0pt]
\item Abundant ideas \& freedom
\item Risk of collapse without structure
\item Race to the bottom dynamics
\item Goal: channel energy into stability
\end{itemize}
\\
\multicolumn{2}{c}{} \\[0.3em]
\multicolumn{2}{r}{\textbf{High Aspiration →}}
\end{tabular}
\end{table}

The target state is Quadrant II (High Potential). Economists Daron Acemoglu and James Robinson describe a similar state as the ``narrow corridor,'' where liberty thrives between the fear of state repression (despotism) and the fear of violence (anarchy) (Acemoglu \& Robinson 2019). In the age of AI, the high stability / high aspiration quadrant is where society reaps the benefits of the Digital Gorilla without being crushed by it, where the State maintains enough centralized control to prevent catastrophe, while Enterprises and People hold enough distributed agency to fuel innovation. Policy interventions should aim to reach or maintain this quadrant: from Quadrant I (Surviving not Thriving), open space for aspiration; from Quadrant III (Creative Chaos), channel energy into stable structures; from Quadrant IV (Fragile Collapse), emergency stabilization is required.

Constitutional democracies are designed to operate in Acemoglu and Robinson's ``narrow corridor'': rights-protecting courts, independent regulators, and competitive markets are institutional devices for keeping the polity in a zone of both liberty and order (Acemoglu \& Robinson 2019; Youngstown v. Sawyer 1952; Hamdi v. Rumsfeld 2004). The Aspiration-Stability Matrix extends this logic to AI governance by asking whether AI-related interventions move the system toward or away from an analogous constitutional ``corridor'', \textit{i.e.}, Quadrant II, where high aspiration is matched by high stability.

\subsection{Operationalizing the Balance}

Operationalizing the federalist architecture described in Section 4.2 and 4.3 requires applying it across all power modalities, enabling the productive tension between aspiration and stability to hold: Is there enough centralization to maintain directional coherence, and simultaneously sufficient pluralism to prevent any single actor from capturing the entire ecosystem (Landemore 2020)?

\begin{enumerate}
\item[(a)] \textbf{Economic:} Centralized strategic investment prevents winner-take-all concentration, while distributed market competition ensures innovation and efficiency (Mazzucato 2013; Rodrik 2014).

\item[(b)] \textbf{Epistemic:} Shared foundational research and data commons (centralized) combined with diverse analytical approaches (distributed) prevent any actor from monopolizing knowledge production (Jasanoff 2004; Kitcher 2011).

\item[(c)] \textbf{Narrative:} Platform governance standards (centralized) alongside diverse media ecosystems (distributed) maintain information flow without enabling censorship or disinformation dominance (Napoli \& Obar 2014; Suzor 2019).

\item[(d)] \textbf{Authoritative:} Core rights and safety protections (centralized) combined with local democratic control (distributed) preserve legitimacy while enabling adaptation.

\item[(e)] \textbf{Physical:} Coordinated control over critical infrastructure (centralized) alongside local security and operational authority (distributed) prevent both fragility and authoritarianism.
\end{enumerate}

In the case of AI, this sort of healthy federated architecture would protect the multi-actor balance that prevents domination by any single actor, a configuration that history suggests is a precondition for prosperity, while ensuring governance serves human ends rather than merely system maintenance (North 1990; Sen 1999; Acemoglu \& Robinson 2012).

To assess whether the federalized architecture remains properly calibrated and society remains in Quadrant II of the Aspiration-Stability Matrix, metrics like safe/unsafe or competitive/uncompetitive must be replaced with dynamic indicators that measure both aspiration and stability continuously and serve as proxies for tracking power allocation among the actors (Mattern \& Zarakol 2016; Farrell \& Newman 2019).

Aspiration measures would track whether the system generates conditions for human flourishing: Can individuals and communities make meaningful choices about AI adoption and use? Are AI's benefits broadly shared or narrowly captured? Does the environment encourage diverse experimentation? Do actors trust each other enough to cooperate toward shared goals? Aspiration metrics could include, for example, startup formation rates, geographic distribution of compute and other benefits, or public trust measures (Putnam 2000, Stiglitz; Sen \& Fitoussi 2010; Nussbaum 2011).

Stability measures would track whether power remains appropriately distributed: Are AI capabilities, data access, and computational resources allocated across the ecosystem rather than monopolized? Can each societal actor effectively check the others' power? Are power shifts gradual enough that actors can adapt, or do sudden shocks destabilize the system (Wu 2018)? Stability indicators could include, for example, market concentration indices, effectiveness of accountability mechanisms, or system volatility during shocks (Gilens \& Page 2014).

These design principles translate into familiar legal doctrinal tools -- though notably, different from those typically used for conventional technologies. Centralized governance mechanisms correspond to constitutional and administrative constraints such as allocations of authority, nondelegation limits, and standards for judicial review of AI-mediated decisions (Hampton v. United States 1928; Motor Vehicle Mfrs. Ass'n v. State Farm 1983). Distributed governance mechanisms map onto competition law, information rights, and individual contestation rights that preserve pluralism in the face of concentrated AI power (Goldberg v. Kelly 1970; Mathews v. Eldridge 1976). The Four Societal Actors model thus does not displace existing fundamental doctrines, but it does reorganize them around the central question of how law should allocate and rebalance power among People, the State, Enterprises, and AI.

For clarity, Quadrant II is not a static endpoint; it requires an ongoing, dynamic game of rebalancing. This would be enabled by continuous monitoring, feedback loops, and trigger conditions that signal when rebalancing is needed before imbalances calcify into permanent concentrations or fragment into collapse (Folke 2005; Walker \& Salt 2012). From a constitutional perspective, this is analogous to periodic review of emergency powers, sunset clauses on delegations of authority, and ongoing judicial and legislative oversight of administrative agencies. AI governance will likewise require institutions empowered to revisit, revise, and, when necessary, revoke AI-related allocations of power as conditions change.

This dynamic approach acknowledges that AI governance is an ongoing institutional challenge. The framework proposed in this article provides both the diagnostic tool (Aspiration-Stability Matrix) and architectural principles (federalism) to navigate this evolution, preserving productive tension between order and opportunity as AI capabilities continue to advance.

A note on execution: theoretical architecture is cleaner than bureaucratic reality. As seen in large organizations attempting federated AI governance models, the risk is governance theater, where committees proliferate but decisions stall. For the Madisonian model to function, the centralized anchors must be lean and high-authority, while the distributed layers must be empowered to move at the speed of code. Poor execution will result in bureaucratic stasis (Quadrant I) rather than Quadrant II. Conversely, overcorrecting toward speed without safeguards risks a slide into Quadrant III or IV. Getting the institutional design right is therefore not just a matter of adopting the right principles on paper, but of embedding them in legal structures, like constitutional provisions, statutes, regulations, and enforceable oversight mechanisms that can survive the pressures of real-world politics and technological change.

\section{Conclusion: Toward an Evolution of Human Governance}

The central argument of this article implies that the challenge of the AI era is not merely technological, but constitutional. As such, AI governance must move from reactive crisis management toward proactive institutional redesign. Just as successful constitutional democracies built lasting institutions by recognizing that power must be balanced rather than concentrated, effective AI governance requires institutionalizing checks and balances that keep all four societal actors in productive tension rather than allowing any one -- including AI itself -- to dominate (Collingridge 1980; Jasanoff 2016).

The Four Societal Actors model recognizes the subset of AI systems that now exhibit distinctive qualities (distributed collective intelligence, normative embeddedness, operational opacity, infrastructural embeddedness) and capabilities (epistemic, decisional, implementation, coordination, and norm-setting), functionally rendering them societal actors. These are not anthropomorphic attributions but observable properties that create political effects as AI systems structure choices, coordinate behavior, and set norms in ways that other actors must accommodate.

Recognizing this reality shifts AI policy from a technology regulation problem to a coordination problem among four actors with competing interests and asymmetric capabilities. It reframes the core questions from technical ones, like: How do we make AI safe? How do we keep AI competitive? to political ones, like: How do we ensure AI's growing epistemic power does not permanently undermine democratic deliberation? What prevents Enterprises and AI systems from co-opting regulatory processes? Under what conditions can People effectively constrain the AI systems they depend on? How do we prevent any single actor from monopolizing power across multiple modalities simultaneously?

\subsection{Research Agenda}

Answering these questions requires an ambitious research agenda that builds on the frameworks and models developed in this article.

\begin{enumerate}
\item[(a)] \textbf{Power dynamics and strategic games.} Section 3 introduced the concept of power modalities and conflict pairs. The framework developed here can be used to deepen analysis of how the four societal actors deploy power across the five modalities in response to one another. Future work may develop detailed mapping of specific conflict pairs (AI $\leftrightarrow$ State, Enterprise $\leftrightarrow$ People, etc.) and trace how power shifts cascade across modalities. By treating these interactions as strategic games rather than static relationships, we can better predict where dangerous concentrations will emerge and which interventions might restore balance. A key empirical question is whether different classes of AI systems (for example, distinct foundation models subjected to divergent post-training regimes) converge toward similar patterns of power deployment or instead give rise to heterogeneous ``AI actor types'' with distinct incentives and institutional footprints (Ouyang et al. 2022; Bai et al. 2022).

\item[(b)] \textbf{Scenario-based validation through case studies.} The Four Societal Actors model can be further tested against both real-world and projected conditions. One promising approach is to forecast future scenarios and use these as case studies for applying and refining the framework. Such scenarios might span the knowledge economy (where AI systems govern global information access), capital markets (where algorithmic trading and credit allocation dominate), and personal productivity and wellness (where AI systems increasingly manage aspects of daily life). By examining how power shifts under different technological trajectories and policy configurations, scholars and policymakers can identify which governance architectures remain robust across multiple plausible futures.

\item[(c)] \textbf{Policy lever library.} This article identifies five power modalities and illustrates how particular policy instruments can rebalance power when problematic concentrations emerge. Future work can extend this analysis by developing a more comprehensive ``menu'' of policy interventions and systematically mapping their effects on specific actor-modality combinations. By cataloging these levers and their interactions, subsequent scholarship can equip policymakers and practitioners across sectors with a practical toolkit for diagnosing imbalances and selecting appropriate interventions.

\item[(d)] \textbf{Dynamic modeling through policy simulation.} One core insight of this work is that AI governance requires adaptive rebalancing as capabilities and contexts evolve. The Four Societal Actors model could be operationalized through simulation environments that model interactions among the four actors under different conditions, allowing policymakers, researchers, and stakeholders to explore how different policy levers might affect system dynamics prior to real-world implementation (Axelrod 1997; Epstein 2006).
\end{enumerate}

Supporting these four tracks, parallel investigations into legitimacy and accountability mechanisms would be valuable for examining how AI's unique authority structure, which is derived from perceived technical objectivity rather than democratic mandate or market validation, creates novel governance challenges that neither communicative repair (characteristic of State legitimacy) nor competitive displacement (characteristic of Enterprise legitimacy) can easily address.

\subsection{Scaling from Local to Global}

While the insights presented in this article draw upon the American federalist tradition, they have the potential to extend beyond individual nations. Just as traditional federal systems balance central and regional powers, global AI governance must balance international coordination with national sovereignty and policy diversity (Jayasuriya 1999).

Through treaties, standards bodies, and multilateral agreements, sufficient coordination on existential risks, basic rights protections, and interoperability can be achieved without demanding global uniformity. Such uniformity would be both politically infeasible and epistemically unwise, eliminating the distributed learning that comes from varied approaches (Sabel \& Zeitlin 2012). A polycentric structure, with multiple centers of authority operating at different scales with overlapping jurisdictions, creates resilience through redundancy while enabling adaptation through diversity (Ostrom 2010; Dafoe 2021). This preserves national and regional autonomy and allows societies to reflect different cultural values while contributing to collective knowledge about what works.

The Four Societal Actors framework provides the analytical foundation for this polycentric approach by making explicit what needs coordination (preventing catastrophic power concentrations that threaten all actors) versus what benefits from diversity (enabling different societies to experiment with varying balances among the four actors). The same actor-modality mapping can also illuminate AI governance challenges and innovation opportunities in non-democratic contexts, for example, authoritarian regimes where the State dominates all modalities, or state-capitalist systems, where Enterprises function as State extensions. Exploring how the framework operates in those settings is an important direction for future work.

\subsection{Implications for Policymakers and Jurisprudence}

For policymakers, the frameworks presented in this article imply several immediate shifts in approach:

\begin{enumerate}
\item[(a)] \textbf{First, stop analogizing AI to existing technologies.} The instinct to map AI onto inherited categories (platform, infrastructure, general-purpose technology) has produced incoherent regulation. Instead, policymakers should ask: Which actor is gaining power? How is value being created, captured, or displaced, and at whose expense? This diagnostic clarity, enabled by the actor-modality mapping and conflict pair analysis, enables more targeted interventions.

\item[(b)] \textbf{Second, monitor power distributions, not just technical capabilities.} Current regulatory approaches focus on AI system characteristics (model size, training data, deployment context). The Four Societal Actors framework suggests monitoring power concentrations across modalities instead. Market concentration indices, epistemic asymmetries, narrative volatility, and authority delegation patterns become the key metrics. The Aspiration-Stability Matrix provides the evaluative framework: Are we preserving both human flourishing (aspiration) and balanced power (stability), or drifting toward authoritarian control, chaotic fragmentation, or failed governance?

\item[(c)] \textbf{Third, design for dynamic rebalancing, not static equilibrium.} Attempting to engineer perfect balance through comprehensive ex ante regulation risks brittle failure. Complex systems with multiple actors resist top-down control. Effective governance requires coordinated interventions across multiple modalities simultaneously, with continuous feedback loops to detect whether desired rebalancing is occurring or whether the system is adapting around the intervention. This means maintaining diverse policy portfolios that can be recalibrated at speed to strengthen weakened actors or constrain overconcentrated ones depending on evolving conditions.

\item[(d)] \textbf{Fourth, institutionalize productive tension.} A federated governance architecture depends on maintaining both centralized anchors and distributed autonomy. Policymakers must resist pressure to resolve this tension in favor of one principle. The tension itself between Aspiration and Stability is what keeps the system in Quadrant II. Making both dimensions explicit and treating them as equally legitimate policy objectives provides a way to evaluate trade-offs systematically rather than ideologically.
\end{enumerate}

For ensuing legal doctrine, the payoff is a reframing rather than a wholesale replacement. Administrative law, constitutional law, and corporate law already contain tools for constraining powerful actors and legitimating the delegation of decision-making authority. What they lack is a coherent picture of AI as one of those actors. By articulating AI's distinctive modalities of power and embedding them within a four-actor constitutional structure, this article provides that picture. Future doctrinal work can build on this foundation to refine standards for delegating authority to AI systems, structure due process protections around AI-mediated decisions, and clarify the fiduciary and corporate responsibilities of entities that deploy AI as de facto decision-makers.

\subsection{Inverting the Intelligence Explosion}

Here we return to Good's ``intelligence explosion'' hypothesis, but inverted: rather than asking whether AI will recursively improve itself beyond human control, we ask whether humans can recursively improve our capacity to harness increasingly capable AI in productive and desirable ways (Good 1966).

The Digital Gorilla is already in the room. The work ahead is not to wrestle it into submission with a set of chains, but to build the institutional environment where its strength can contribute to a stable society where human aspiration can blossom. In that sense, AI governance is less a technical control problem than an exercise in constitutional imagination: a test of our capacity to invent institutions that can live with, and make use of, non-human forms of agency. This requires moving AI governance from reactive crisis management toward proactive institutional design: creating the checks and balances, the monitoring systems, the policy levers, and the adaptive capacity that allow all four societal actors to constrain each other's power while enabling productive collaboration.

The frameworks provided in this article lay the conceptual foundation; the research agenda outlined above sketches how it can be extended through deeper analysis of power dynamics and strategic games, scenario-based validation through case studies, richer policy lever library, and interactive simulation environments that build institutional capacity for anticipatory governance. The task that follows is to embed this architecture into the institutions capable of sustaining balanced, aspirational AI governance.

If we succeed, the defining characteristic of the AI age will not be the obsolescence of human governance, but its evolution.

\section*{Acknowledgments}
M. Alejandra Parra-Orlandoni gratefully acknowledges Ruxanda Renita, M.P.A. candidate, Harvard Kennedy School, and Ignacio Urreo Bordones, Visiting Fellow, Harvard Economics Department, for outstanding research assistance and contributions to this article. For valuable comments and conversations, she thanks Richard Zeckhauser, Frank Plumpton Ramsey Professor of Political Economy, Harvard Kennedy School; Martha Minow, 300th Anniversary University Professor, Harvard University; John Haigh, Co-Director, M-RCBG, and Lecturer in Public Policy, Harvard Kennedy School; Dan Murphy, Executive Director, M-RCBG, Harvard Kennedy School; Steve Weber, Partner, Breakwater Strategy, and former Professor of Political Science and founding Director of the Center for Long-Term Cybersecurity, University of California, Berkeley; Jorge Kamine, Partner, Willkie Farr \& Gallagher LLP; and the extraordinary M-RCBG 2025-26 cohort. In particular, she thanks Dan Levy, Senior Lecturer in Public Policy, Harvard Kennedy School, for his guidance as her advisor on this project. She is especially indebted to Aaron Mauck, her husband and former Faculty of Arts and Sciences at Harvard University, for his patience with my many questions and for his generous advice.

\end{document}